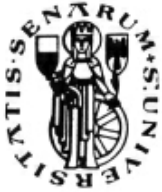

# UNIVERSITÀ DEGLI STUDI DI SIENA

# Dipartimento di Economia Politica

# Interlocking editorship. A network analysis of the links between economic journals

*Alberto Baccini\* and  Lucio Barabesi°*

\* Dipartimento di Economia Politica, Università di Siena (baccini@unisi.it)
° Dipartimento di metodi quantitativi, Università di Siena (barabesi@unisi.it)

ABSTRACT: The exploratory analysis developed in this paper relies on the hypothesis that each editor possesses some power in the definition of the editorial policy of her journal. Consequently if the same scholar sits on the board of editors of two journals, those journals could have some common elements in their editorial policies. The proximity of the editorial policies of two scientific journals can be assessed by the number of common editors sitting on their  boards. A database of all editors of ECONLIT journals is used. The structure of the network generated by interlocking editorship is explored by applying the instruments of network analysis. Evidences have been found of a compact network containing different components. This is interpreted as the result of a plurality of perspectives about the appropriate methods for the investigation of problems and the construction of theories within the domain of economics.

Research funded by Ministero dell'Università e della Ricerca PRIN 2005 "The evaluation of economic research in a historical perspective: comparing methods and arguments"

Draft version 14/04/2008 r.1.1

COMMENTS WELCOME

The aim of this paper is to explore the structural properties of the network generated by the editorial activities of economists. The domain of the research is the academic community of the economists involved in journals included in the *Econlit* database maintained by the American Economic Association. The basic intuition of our research is that studying the structure of the network of the economic journals, with the instruments of network analysis, can shed some light on the underlying processes according to which the research is conducted by economists.

More in general this intuition is based on the common experience that scientific activities have a not reducible social dimension (Longino, 2006 and the bibliography cited therein). It is usual to ask a colleague to collaborate in writing a paper, to comment a book, to revise a project. And it is usual when judging the quality of a paper, a research outcome or of a research project to commit it to the opinions of experts or *peers*. The editorial boards of scientific journals decide which papers are worthy of publication by asking the opinion of anonymous referees. The proxies normally used for measuring the scientific quality of a paper or of a journal, e.g. the well known Impact Factor (IF), are implicitly based on the relational dimension of scientific activity. The bibliometric popularity indeed depends on the number of the citations that the articles of a scholar or of a journal receives by articles authored by other scholars principally in the same domain of research. In some cases, the relevance of individual scientific activity is approximated by indicators of the *esteem*. The esteem is based on the positive appraisal that other scholars give to an individual, and this positive appraisal is reflected in the position he occupies in the scientific community, as, for example, director of a research project, of a scientific journal and so on.

All of the agents and actions described above can be viewed as interdependent rather than as autonomous units; the actions can then be considered as relational ties (linkages) between agents (Wasserman and Faust, 1994). The patterns of connections between agents form a social network, and the structure of such networks affects the social interactions amongst agents (Goyal, 2007; Newman, In press). Such a network can be represented as a set of points (the so called nodes) denoting actors, joined in pairs by lines (the so-called edges) denoting acquaintance. The quantitative empirical studies of such networks can be conducted with the instruments of network analysis (Wasserman and Faust, 1994).

The network analysis was already applied in the study of network generated by scientific activities. The most frequented topic is that of scientific collaboration. In this case, two scientists are considered connected if they have co-authored a paper together. Newman (2000; 2004) analyzed the collaboration networks of scientists from biology, medicine and physics and found that all networks constitute a "small world" (Barabási, 2003) in which the average distance between scientists via intermediate collaborators is very small. Goyal, Van der Leij and Moraga-González (2004) studied the coauthorship network in economics and found evidences that in the period 1970-2000 the network became a small world. This feature was explained in reference to a decision model according to which authors rationally decide whether to work alone or with some co-authors (Fafchamps, Van der Leij and Goyal, 2006).

Another application of network analysis to the study of scientific activity concerns the network generated by citations either in patents (Jaffe and Trajtenberg, 1996) or scientific paper (Otte and Rousseau, 2002). In this latter case the vertices of the network are papers and a (direct) link is a citation of a paper by another paper. It is easy to see that hyperlinks in the WWW are a general case of a citation network. The networks generated by the citations of scientific papers and by WWW hyperlinks show a strong characterization in terms of the degree distribution of the nodes. The degree of a vertex is the number of links with the other vertices in the network. The degree distribution strongly characterizes the underlying network pattern. In the above cited as in many other cases, empirical investigations have shown that the degree distributions often display Paretian tail behaviour (see e.g. Newman, 2005 and the reference therein). The probability to be cited can be affected by the number of citation already collected by a paper (Simkin and Roychowdhury, 2006). This self reinforcing mechanism can push the authors to preferential attachment, i.e. in this case the tendency among papers in the growing network of citations to form new links citing preferentially to papers with a high number of citations and, as a consequence, to the emergence of star papers and authors (Barabási and Albert, 1999; Faria, 2005).

This paper proposes a new application of the standard network analysis techniques. We focus on the network of economic journals generated by the presence of at least a same editor in their boards. The vertices of the network considered are scientific journals, and a (not directed) link is generated between a pair of journals by the presence of at least a same editor in the board of both. This network is generated by a simple transformation of the dual-mode or affiliation network (de Nooy, Mrvar and Batagelj, 2005; Wasserman and Faust, 1994). An affiliation network is a network in which the vertices are divided in two sets (actors and events) and the affiliation connects vertices from the two different sets only. In our case affiliation (being member of the board) connects a scientist to a journal. Our attention will be focused on the links generated between journals through the affiliations of scientists. It is worth remarking that the present framework is similar to that considered in interlocking directorship analysis, which is probably the most developed field of application of dual mode network analysis. An interlocking directorate occurs when a person sitting on the board of directors of a firm also sits on the board of another firm. Those interlocks have become the primary indicator of inter-firm network ties. An inter-firm tie can be explained as the result of a strategic decision of the firms, as collusion, or cooptation or monitoring of sources of environmental uncertainty (Mizruchi, 1996 ant the reference cited therein). In analogy with interlocking directorates analysis, we will refer to the phenomenon here considered as *interlocking editorship*.

The issues on which we focus are: which are the most central journals of the network and which are the most peripheral? Which journals have the most influence over others? Does the community of economists break down into smaller groups? If so what are they? More in general is it possible to separate schools of thought, methodologies or pattern of research characterizing the scientific community under scrutiny? And it is possible to infer something about the functioning of the "market of research" in the domain of economics?

## *Editorial boards*

Cogent answers to those questions require a significant unit of observation at the basis of our research. To the best of our knowledge, no literature presents extensive discussions about the role of the board of editors for scientific journals; but we have anecdotal evidence and some recent tentative generalizations. Traditionally the main function of the board of editors was to determine which articles were appropriate for publication. In the last two or three decades this function has changed. The diffusion of the anonymous referees process allows the board of editors to work for obtaining and evaluating referees: "The job of both the editors and the referees is to do 'peer review'" (Coupe, 2004). In every case the role of editors can be considered of some relevance in steering discipline, and pushing or suppressing various lines of research (for some anecdotal evidence see Stigler, Stigler and Friedland, 1995 and the reference therein). Probably an editor's objective is to produce a journal of high quality. The advent of modern bibliometric indicators, as the impact factor, slightly modified the traditional problem of editors: journal editors compete now with each other to attract the *best* papers, i.e. the papers with highest probability to be cited. The instruments used by editors to attract the *best* papers are the reputation of the journal, the reduction in transaction costs involved in the publication process, but also favoritism for individuals or institutions (Laband and Piette, 1994; Medoff, 2003).

Recently some general models of the functioning of the market for publication has been developed. Faria (2005) models the market for research as a game populated by two kind of agents: authors and editors, with some market power. Authors seek to maximise the number of their publications and the impact of their work in the literature, captured by citations. Editors maximise the quality of papers they publish in order to increase the reputation of their journals. High reputation journals are journals often cited in the literature, i.e. journals with high impact factor. Authors searching to increase the probability to be cited, compete to publish in journals with a high reputation. As a consequence "editors of journals with strong reputations enjoy an enormous amount of power in their hands" (Faria, 2005). The competition between authors generates a pressure to publish which open the space to new journal and new editors, which, at their turn, seek to position their journals in empty niches and to adopt strategies to improve their impact factor. The overall result of this mechanism is largely positive driving the system toward the enhancement of research quality (Goel and Faria, forthcoming). A more critical view is developed by Frey (2003; 2005) starting from a property rights approach. According to Frey, there are more than two groups of actors in the academic publishing system: notably publishers, editors, referees and authors. Each groups owns different property rights on the scientific journal. And the behavior of each group must be modeled according the traditional rational choice model of man. Commercial publishers owning complete private property rights on various journals, search to maximize profits. They are interested in the quality and reputation of the journals. A high quality journal has market power, and market power permits the gain of profits. Editors enjoy property rights on their journal; they are interested in the quality of their journal because this enhances their academic reputation. Authors, as we have seen in Faria's model, are interested in publishing their article in high reputation journals. A large amount of decision power over the publishing process is detained by the referees who have no property right at all on the journals. So it is at least incoherent to model them as acting only in the interest of the science or quality of the research, as for example in Engers and Gans (1998). Personal interest can play a role in their choices. "Many referees will be tempted to judge papers according to whether their own contribution are sufficiently appreciated and their own publications quoted" (Frey, 2003). The diffusion of anonymous referee processes reduced drastically, at least in economics, the power of editors. This reduction can have negative consequences on the overall quality of research, driving the authors to "intellectually prostitute themselves by slavishly following the demands made by anonymous referees who have no property rights to the journals they advise" (Frey, 2003). Evidence on three of the most eminent journals in economics shows that in the last years the power of editors in accepting or rejecting paper is anew growing, driving, as a consequence, to the publication of papers authored by scholars concentrated in few highly reputed institutions in the United States (Wu, 2007).

From the point of view of this article, the moral of the story is trivial: the board of editors have some power in shaping the editorial processes and policies of scientific journals. Because of their importance, it appears reasonable that the positions on the boards are held by "persons who have the confidence and trust of their colleagues in the journal's areas of coverage for the journal to be successful in attracting quality submissions and in building and maintaining a reputation for quality. Thus, selection as an editor or member of an editorial board is a considerable honor that reflects one's standing in the profession as evaluated by his or her peers" (Kaufman, 1984). Unsurprisingly, members of the editorial board usually place this information in evidence in their *curriculum vitae*; and it is universally recognized that to be a member of an editorial board is a signal of the esteem reserved to a scholar by the academic community.

The esteem or reputation of a scientist is shared also by her institution given that her affiliation is clearly displayed in the journal. This point was stressed by some empirical papers which ranked academic institutions on the basis of their representation on the editorial boards of top journals (Gibbons and Fish, 1991; Kaufman, 1984). Interestingly enough, Gibbons and Fish (1991) found that there is a correlation between the ranking of institutions based on the membership of their affiliated scholars on editorial boards and those based on citations and publications in top journals. Hodgson and Rothman (1999) did not find surprising at all that editors and authors who ranked according to some quality (merit, prestige, citation) came from the best institutions ranked according to the same criterion. This can be explicable in terms of "familiar mechanism and cultural processes" (Hodgson and Rothman, 1999). In a nutshell: "having an editor increases the member of the departments' chance to publish in that journal" (Coupe, 2004 and the references cited therein). In the case of economics, Hodgson and Rothman present data evidencing a strong concentration of editors and authors in few academic institutions of the United States, and argue that this institutional oligopoly may be unhealthy for the development of innovative research.

As already anticipated, this paper approaches the question of the membership of an editorial board from a perspective slightly different from the ones seen above. We are interested in the relation between the editorial policies of the various journal and we would infer some considerations on this topic by studying the cross-presence of the editors in their boards.

It is apparent that editorial boards have some power in shaping the editorial processes and policies of economic journals. Therefore, our perspective is based on the hypothesis that each editor may influence the editorial policy of her journal. Consequently if the same individual sits in the board of two journals, those journals could have some common elements in their editorial policies. It is evident that we will not be concerned with direct observations of the editorial policies adopted by the boards of economic journals. We will infer considerations about the similarity of editorial policies through the observation of the presence of scholars in the boards of editor.

*Centre and periphery in the interlocking editorship network*

First, it should be emphasized that the empirical notion of editor adopted in this paper is very broad. Indeed, it covers all the individuals listed as editor, co-editor, member of the editorial board or of the advisory editorial board. There is no evidence regarding the roles of different kinds of editors in the editorial process (possibly apart from the role of editor-in-chief) and a single title such as managing editor may often entail very different roles for different journals. Hence, as in Hodgson and Rothman (1999), the broad definition is assumed.

The affiliation network database was constructed *ad hoc* for this paper. We have included in our research 746 journals present in the ECONLIT database and with an active editorial board in January 2006. That is we excluded from the database journals ceased before January 2006. According to common wisdom, this set of journals includes all major scientific journals in the field of economics.

The data on the members of the editorial boards was directly obtained from the website of the journals or - for the few cases when the site was unavailable - from the hard copy. The data was collected from March to July 2006 considering the boards published on the websites of the journals in that period. When the hard copy was necessary, the board considered was that of the first issue in 2006 or, alternatively, that of the last issue in 2005. Moreover, the database was managed by means of the package *Pajek* (Batagelj and Mrvar, 2006; de Nooy et al., 2005).

In this database, 21,525 seats were available on the editorial boards and they were occupied by 15,921 scholars. The average number of seats per journal turned out to be 28.9, while the average number of seats occupied by each scholar (*i.e.* the mean rate of participation) was 1.35. The number of lines linking the journals is 6,407, and the density of the interlocking directorship network (*i.e.* the ratio of the actual number of lines to the maximum possible number of lines in the network) is 0.023. This means that only 2.3% of the possible lines is present (Wasserman and Faust, 1994). A comparison with statistical journals network (Baccini, Barabesi and Marcheselli, 2008) shows that economic journals network is much more dispersed than the statistical one.

The graph of the network is reported in Figure 1. The vertices in the graph are automatically placed by the package *Pajek* on the basis of the Fruchterman-Reingold procedure. In this graph two main subsets may be roughly recognized: a giant central component composed by the majority of economic journals and a small group of isolated journals.

**Figure 1.** The economic journals network.

In order to consider an initial exploratory analysis, the degree distribution has been provided. In the present setting, the degree of a journal is the number of lines which it shares with the other journals. Table I contains the degree distribution of the journals considered. The mean degree is 17.18 (while the median degree turns out to be 11) and the degree standard deviation is 17.55. It is interesting to remark that about 10% of journals, precisely 74 journals, are isolated from the network (*i.e.* they have zero degree). They are in part non-English language journals (as for example *L'impresa* or *Bancaria* in Italian, *Tahqiqat-e eqtesadi* in Arabian, *Investigación Económica* in Spanish), journals edited by national scientific societies (e.g. *Schweizerische Zeitschrift für Volkswirtschaft und Statistik/Swiss Journal of Economics and Statistics* edited by the Swiss Society of Economics and Statistics) or by institutions with a complete control of the board of editors (e.g. the *Antitrust Bulletin*) journals dedicated to very narrow topics (e.g. *Australian Commodities Forecasts and Issues* or *Agronomia Mesoamericana* ; *Australian Bulletin of Labour*); journals on the boundaries with other disciplines with only a minor emphasis in economics (e.g. *American Historical Review* or *Transportation Journal*). Indeed the search for components in the network (de Nooy et al., 2005) trivially shows 74 components each made up of one element (the aforementioned journals), a component made up of two journals (both edited by the American Statistical Society: *American Statistician* and the *Journal of the American Statistical Association*), and a big component made up of the remaining 670 journals. As usual, a component is a maximal connected sub-network, *i.e.* each pair of sub-network vertices are connected by a sequence of distinct lines (de Nooy et al., 2005)

**Table I.** Degree frequency distribution of the statistical journals.

| Degree | Freq | Freq% | Degree | Freq | Freq% | Degree | Freq | Freq% |
|---|---|---|---|---|---|---|---|---|
| 0 | 74 | 9,9 | 24 | 8 | 1,1 | 48 | 5 | 0,7 |
| 1 | 44 | 5,9 | 25 | 6 | 0,8 | 49 | 1 | 0,1 |
| 2 | 39 | 5,2 | 26 | 13 | 1,7 | 50 | 3 | 0,4 |
| 3 | 37 | 5,0 | 27 | 13 | 1,7 | 51 | 4 | 0,5 |
| 4 | 23 | 3,1 | 28 | 11 | 1,5 | 52 | 1 | 0,1 |
| 5 | 32 | 4,3 | 29 | 10 | 1,3 | 53 | 3 | 0,4 |
| 6 | 19 | 2,5 | 30 | 7 | 0,9 | 54 | 4 | 0,5 |
| 7 | 25 | 3,4 | 31 | 3 | 0,4 | 55 | 2 | 0,3 |
| 8 | 21 | 2,8 | 32 | 9 | 1,2 | 57 | 5 | 0,7 |
| 9 | 21 | 2,8 | 33 | 5 | 0,7 | 59 | 2 | 0,3 |
| 10 | 19 | 2,5 | 34 | 6 | 0,8 | 60 | 2 | 0,3 |
| 11 | 20 | 2,7 | 35 | 6 | 0,8 | 61 | 1 | 0,1 |
| 12 | 16 | 2,1 | 36 | 7 | 0,9 | 62 | 2 | 0,3 |
| 13 | 22 | 2,9 | 37 | 2 | 0,3 | 63 | 1 | 0,1 |
| 14 | 16 | 2,1 | 38 | 8 | 1,1 | 65 | 2 | 0,3 |
| 15 | 19 | 2,5 | 39 | 6 | 0,8 | 66 | 1 | 0,1 |
| 16 | 12 | 1,6 | 40 | 6 | 0,8 | 68 | 1 | 0,1 |
| 17 | 8 | 1,1 | 41 | 7 | 0,9 | 69 | 2 | 0,3 |
| 18 | 9 | 1,2 | 42 | 5 | 0,7 | 71 | 1 | 0,1 |
| 19 | 12 | 1,6 | 43 | 6 | 0,8 | 72 | 3 | 0,4 |
| 20 | 14 | 1,9 | 44 | 7 | 0,9 | 73 | 1 | 0,1 |
| 21 | 10 | 1,3 | 45 | 3 | 0,4 | 76 | 1 | 0,1 |
| 22 | 11 | 1,5 | 46 | 5 | 0,7 | 79 | 1 | 0,1 |
| 23 | 8 | 1,1 | 47 | 5 | 0,7 | 94 | 1 | 0,1 |
| | | | | | | 124 | 1 | 0,1 |

A main concern in network analysis is to distinguish between the centre and the periphery of the network. In our case, the problem is to distinguish between the economic journals which have a central position in the network and those in the periphery. As suggested by Wasserman and Faust (1994), three centrality measures for each journal in the network may be adopted. The simplest measure for the centrality of a journal is represented by its degree: indeed, the more ties a journal has to other journals, the more central is its position in the network. For example, the *Pacific Economic Review* is linked with 124 journals, while *Journal of Development and Economic Policies* is linked with solely one. Hence, the first is more central in the network than the second. In addition, the normalized degree of a journal is the ratio of its degree to the maximum possible degree (*i.e.* the number of journals minus 1). Thus, the *Pacific Economic Review* is linked with about 16.6% of the other journals in the network, while *Statistical Modelling* is linked with only 0.001%. Table A1 contains the degree and the normalized degree for the journals considered. An overall measure of centralization in the network (based on marginal degrees) is given by the so-called degree centralization (Wasserman and Faust, 1994). In this case, the index turns out to be 0.14, showing that the network of economic journals is less centralized than the one of statistical journals for which the degree centralization is 0.34 (Baccini et al., 2008).

The second centrality measure is given by closeness centrality, which is based on the distance between a journal and all the other journals. In the network analysis, the distance between two vertices is usually based on so-called geodesic distance. Geodesic is the shortest path between two vertices, while its length is the number of lines in the geodesic ((Wasserman and Faust, 1994). Hence, the closeness centrality of a journal is the number of journals (linked to this journal by a path) divided by the sum of all the distances (between the journal and the linked journals). The basic idea is that a journal is central if its board can quickly interact with all the other boards. Journals occupying a central location with respect to closeness can be very effective in communicating information (sharing research, sharing papers, deciding editorial policies) to other journals. Table A1 contains the closeness centrality for economic journals. By focussing on the connected network of 640 journals, it is possible to compute the overall closeness centrality of journals (Wasserman and Faust, 1994). The overall closeness centrality is 0.29, showing in turn that the network of economic journals is less centralized than the statistical one [0.35].

The third considered measure is the so-called betweenness centrality. The idea behind the index is that similar editorial aims between two non-adjacent journals might depend on other journals in the network, especially on those journals lying on the paths between the two. The other journals potentially might have some control over the interaction between two non-adjacent journals. Hence, a journal is more central in this respect if it is an important intermediary in links between other journals. From a formal perspective, the betweenness centrality of a journal is the proportion of all paths between pairs of other journals that include this journal. Table A1 contains the betweenness centrality of the economic journals. For example, the *Pacific Economic Review* is in about 4% of the paths linking all other journals in the network. It is interesting to note that in the statistical journal network, the two journals with higher betweenness are each in about 12% of the paths linking all other journals (Baccini et al., 2008). In turn, the overall betweenness centralization of the network (Wasserman and Faust, 1994) is 0.04; also in the case of this index the centralization is lower than in the network of statistical journal [0.10].

It is worth noting the ranking similarity of the three centrality measures. This item is emphasized by the high value of Kendall's concordance index which equals 0.95 (for more details on Kendall's tau and concordance indexes see e.g.Gibbons and Chakraborti, 1992).

## *Valued network analysis*

It is interesting to consider the strength of the relationship between journals. The network of journals can be characterized as a valued network. More precisely, in a valued network the lines have a value indicating the strength of the tie linking two vertices (Wasserman and Faust, 1994). In our case the value of the line is the number of editors sitting on the board of the two journals linked by that line.

Table III shows the distribution of line values: 74.6% of the links are generated by journals sharing only one editor and about 94% are generated by journals sharing three or less editors.

In social network analysis it is usual to consider lines with higher value to be more important since they are less personal and more institutional (de Nooy *et al.*, 2005). In the case of the journal network, the basic idea is very simple: the editorial proximity between two journals can be measured by observing the degree of overlap among their boards. Two journals with no common editors have no editorial relationship. With an example: the *American Economic Review* and the *Australian Bulletin of Labour* have no common editors, so that their editorial policies can be considered independent of each other.

The opposite situation occurs when two journals have the same board; probably they have a common or, at least shared, editorial policy, i.e. they are *companion* journals. As an example, *Applied Economics* and *Applied Economics Letters* share all their 23 editors. In its "aims and scope" declaration for 2007, the latter explicitly stated that it is the "companion journal" of the former. And again, *Politická Ekonomie* and *Prague Economic Papers* not only share all their 40 editors, but they are published by the University of Economics of Prague, and in their web sites they present the same description of the mission of both journals. In economics, there are a few journals that can be considered properly *companion journals* sharing all their editorial board members. The most common situation is the intermediate one in which two journals share only a part of their board members.

**Table III.** Line multiplicity frequency distribution.

| Line value | Freq | Freq (%) |
|---|---|---|
| 1 | 4780 | 74,61 |
| 2 | 934 | 14,58 |
| 3 | 297 | 4,64 |
| 4 | 145 | 2,26 |
| 5 | 89 | 1,39 |
| 6 | 51 | 0,80 |
| 7 | 33 | 0,52 |
| 8 | 24 | 0,37 |
| 9 | 15 | 0,23 |
| 10 | 10 | 0,16 |
| 11 | 8 | 0,12 |
| 12 | 6 | 0,09 |
| 13 | 3 | 0,05 |
| 14 | 2 | 0,03 |
| 15 | 1 | 0,02 |
| 16 | 4 | 0,06 |
| 19 | 1 | 0,02 |
| 20 | 1 | 0,02 |
| 23 | 1 | 0,02 |
| 24 | 1 | 0,02 |
| 40 | 1 | 0,02 |

Starting from this basis it is possible to define cohesive subgroups, *i.e.* subsets of journals among which there are relatively strong ties. In a valued network a cohesive subgroup is a subset of vertices among which ties have a value higher than a given threshold. In our case, a cohesive subgroup of journals is a set of journals sharing a number of editors equal or higher than the threshold. In our interpretation, a cohesive subgroup of journals is a subgroup with a similar editorial policy, belonging to the same subfield of the discipline or sharing a common methodological approach. Following de Nooy et al. (2005), cohesive subgroups are identified as weak components in *m*-slices, *i.e.* subsets for which the threshold value is at least *m*.

As previously remarked, the network of statistical journals is not compact: there is a big component of 670 journals and all the others are isolated. The search for cohesive subgroups strengthens this path: fixing a minimum value of threshold to $m = 2$ the big component reduces to 474 journals, 13 components emerge of 2-4 journals, and the isolated journals grow to 242. With $m = 3$ the big component reduces to 284 journals and isolated journals grow to 369. With higher threshold value, the network gives rise to components worthy of being noticed here.

In particular we focused our attention on the weak components emerging in 6-slices network. It is possible to isolate 41 components including 176 journals. We comment, first, on the three weak components with the biggest number of journals

Figure 2 contains the representation of the central and biggest component of the network. The 36 journals in this subset of the network have at least 6 common editors. The dimension of each vertex represents the betweenness centrality of the corresponding journal in the complete network. The centre of this component is occupied by the *Journal of Money Credit and Banking*. It is linked directly with 8 journals. Four out of eight have not other links (*American Economic Review*, *Journal of Monetary Economics*, *Journal of Macroeconomics* and *Federal Reserve Bank of New York Economic Policy Review*) and therefore they configure themselves as an efficient star at the centre of the network; the other four out of eight journals bridge the central star to four groups of journals. In the upper right of the figure, *Macroeconomics Dynamics*, is the bridge toward journal of macroeconomic dynamics and computational economics at the boundaries of macroeconomics;[1] on the right *The International Journal of Finance and Economics* is the bridge with a small group of other policy oriented and accessible to non-specialists journals. On the lower left *The Journal of Financial Intermediation* and *The Journal of Financial Services Research* are the bridge toward a group of financial journals; in this group the *European Financial Management* connects also a group of business and marketing journals. On the upper left the *Review of International Economics* is the bridge with a group of journals of international economics and development. So, the central component of the network contains journals of macroeconomics, monetary economics, international economics, a few journals of financial economics and the *American Economic Review* considered by all rankings the most important journal of general economics. It is interesting to note that the central star contains all the journals classified by (CNRS, 2007) as the most important journals of macroeconomics, international and monetary economics. This configuration is probably the outcome of the general consensus achieved in monetary policy (and also in macroeconomics?) by scientists and practitioners, as discussed by Goodfriend (2007).

---

[1] In the aims and scope of *Netmomics* it is stated that "the journal also explores the emerging network-based, real-time macroeconomy with its own set of economic characteristics."

**Figure 2 about here**

**Figure 2.** The central weak component in 6-slices network: macroeconomic, monetary and international economics journals (the dimension of vertices is proportional to betweenness centrality).

Figure 3 contains a second weak component with 12 journals devoted to economic theory, econometrics, game and decision theory. The centre of the component is *Games and Economic Behavior*. It is linked directly to seven journals devoted to the study of mathematical and quantitative methods (*Econometrica, Journal of Mathematical Economics*, *International Journal of Game Theory*, *Journal of Economic Theory*, *Review of Economic Design*), of theoretical public economics (*Social Choice and Welfare*), and experimental economics (*Experimental Economics*). In this case the network is not configured as a star, because there are direct links between some of the seven journals around the central one. In this case, as for the macroeconomic component, all linked journals are in the first or second class in the CNRS ranking (CNRS, 2007). It is useful to note that the *Journal of Economic Behavior and Organization* presents a relatively high betweenness centrality, indicating that this controls the links of the component with the rest of the network of economic journals.

**Figure 3.** A weak component in 6-slices network: economic theory, econometrics, game and decision theory journals (the dimension of vertices is proportional to betweenness centrality).

The third weak component is drawn in Figure 4. It contains journals devoted to urban, spatial and geographical economics, and to real estate economics. At the centre of the component there is a pair of journals, *The Journal of Urban Economics* and *The Journal of Regional Science.* The first is linked through *The Journal of Regional Science* to other journals of geographical economics; the second to journals of housing economics and real estate economics and finance. The journals on the right of the Figure 3 are at the boundaries of economics, as for example the *Journal of Real Estate Literature* which is a general publication of the American Real Estate Society; but they are also relatively isolated in the network of the economic journals, as we can infer by their relatively low betweenness centrality values. The journals on the left of the Figure 3 are more central in the network and they have also an interesting position in the CNRS ranking.

**Figure 4.** A weak component in 6-slices network: urban and regional economics journals (the dimension of vertices is proportional to betweenness centrality).

The other eight weak components of network containing more than three journals are drawn in Figure 5. The first component in clockwise contains five journals dedicated to insurance. The second is the component containing six journals of accounting research. In particular four journals out of six are linked in a complete subnetwork (*Accounting Review, Journal of Accounting Research*, *Journal of Accounting and Economics*, *Review of Accounting Studies*); these four journals are all classified as the most important journals in the field of accounting by the CNRS ranking (CNRS, 2007). The third group contains five journals of environmental economics; two out of five, *Journal of Environmental Economic and Management* and *Ecological Economics* are top ranked in CNRS (CNRS, 2007). On the lower part of the figure there is a line network of five journals of applied finance; and another line network of four journals of finance. It is interesting to note that the journals classified by CNRS as "Finance and Insurance" when analyzed with our technique split in three specialized groups. On the left a component is drawn containing six journals of public economics: in this component there are three highly ranked journals by CNRS (*Journal of Public Economics*; *International Tax and Public Finance* and *National Tax Journal*) and three journals published in Germany. The public choice approach to public economics defines a weak component of three journals (*Public Choice, European Journal of Political Economy* and *Constitutional Political Economy*) presented in Figure 6. The last two components of Figure 5 are strongly characterized for their methodological approach. On the upper left there are six journals sharing an Austrian perspective on the study of political economy and political science. In the centre of the figure there is a component containing journals strongly characterized for the evolutionary approach to the analysis of economics, industrial organization and technological change.

**Figure 5 about here**

**Figure 5.** Weak components in 6-slices network with more than three journals (the dimension of vertices is proportional to betweenness centrality)

Figure 6 contains the weak components with three journals. Again in clockwise, on the right there is a group of law and economics journals; then a group of business history journals; the already mentioned group of public choice journal; three journals devoted to the study of the economics of new technology; a component containing three review of development published by Oxford University; and finally three Brazilian economic journals.

**Figure 6 about here**

**Figure 6.** Weak components in 6-slices network with three journals (the dimension of vertices is proportional to betweenness centrality)

## *Conclusive remarks*

The exploratory analysis developed in this paper relies on a weak hypothesis: each editor possesses some power in the definition of the editorial policy of her journal. Consequently, if the same scholar sits on the board of two journals, those journals could have some common elements in their editorial policies. The proximity of the editorial policies of two scientific journals can be assessed by the number of common editors sitting on their boards. On the basis of this statement, applying the instruments of network analysis, a simple interpretation of the economic journal network has been given.

The network generated by interlocking editorship is at first sight very compact, given that about 90% of the journals considered are linked directly or indirectly forming a compact subnetwork. According to our hypothesis, if the degree of overlapping of editorial boards is considered to explore the editorial policies of journals, it is possible to individuate a lot of different groups of journals. This is probably the result of different perspectives about the appropriate methods for the investigation of problems and the constructing of theories within the domain of economics. The competing visions or approaches to economic research prompt scholars to endorse different languages and visions about the correct view of how to conduct research.

**Figure 1.** The economic journals network

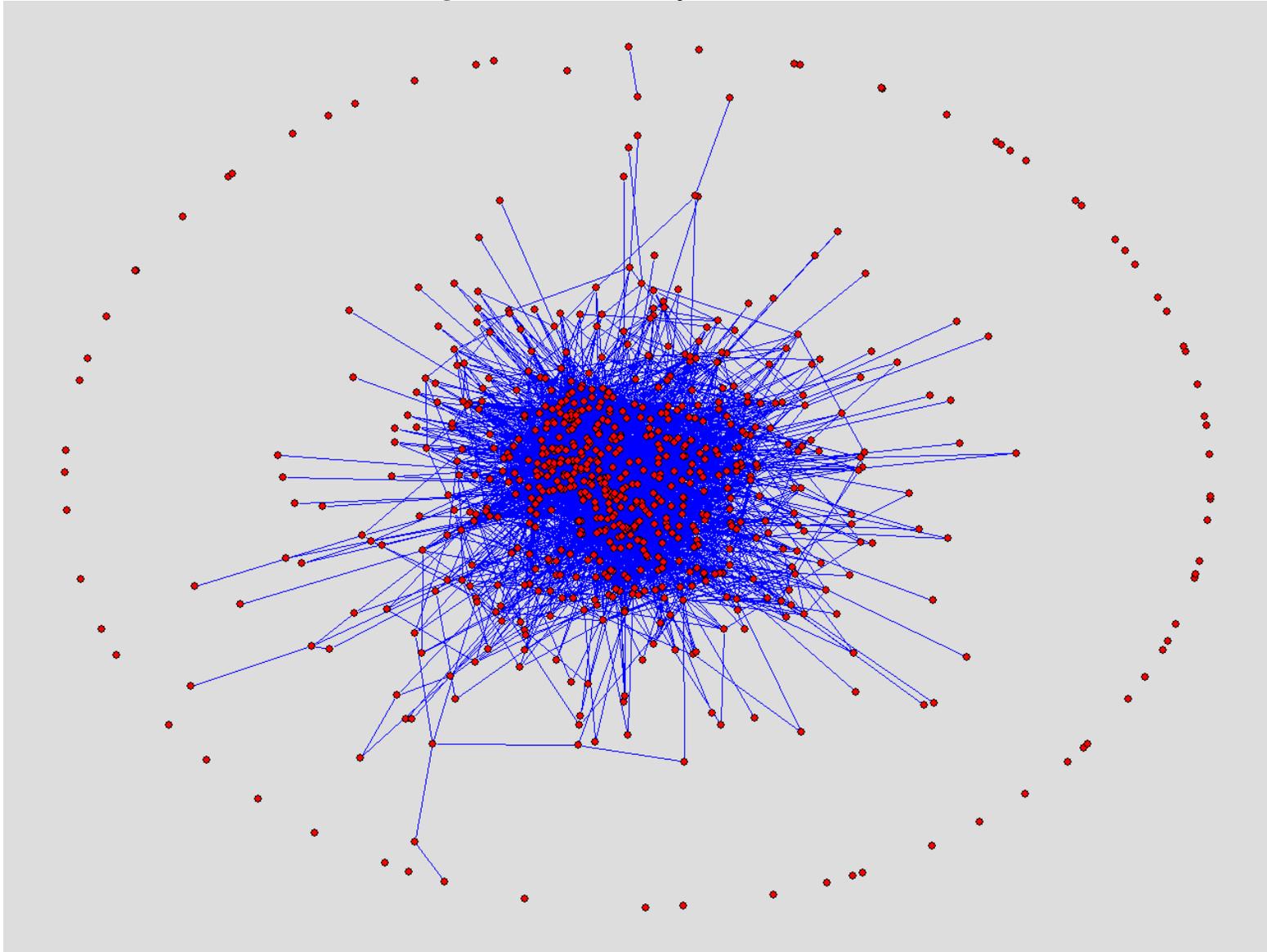

**Figure 2.** The central weak component in 6-slices network: macroeconomic, monetary and international economics journals.

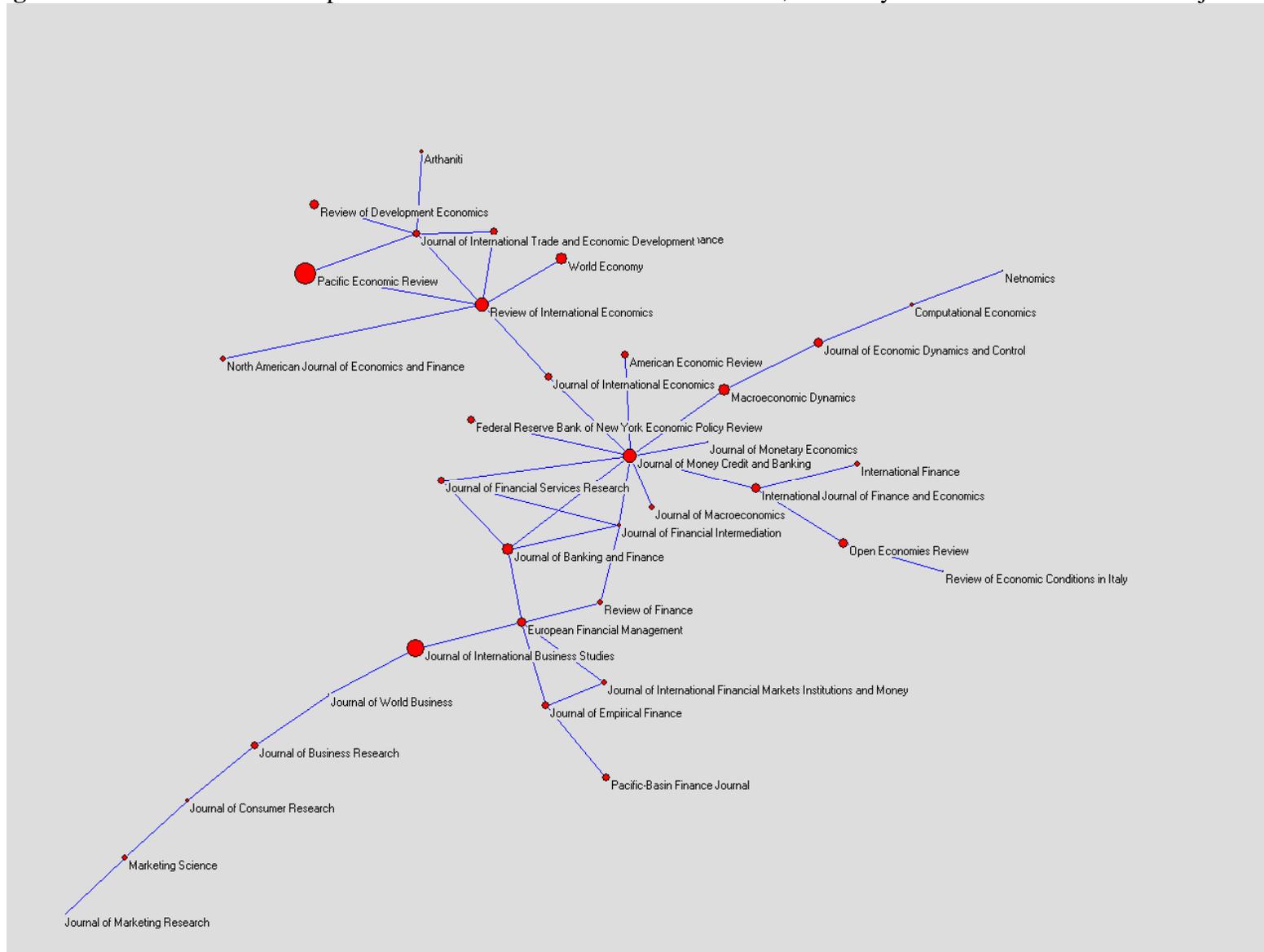

**Figure 3.** A weak component in 6-slices network: economic theory, econometrics, game and decision theory journals

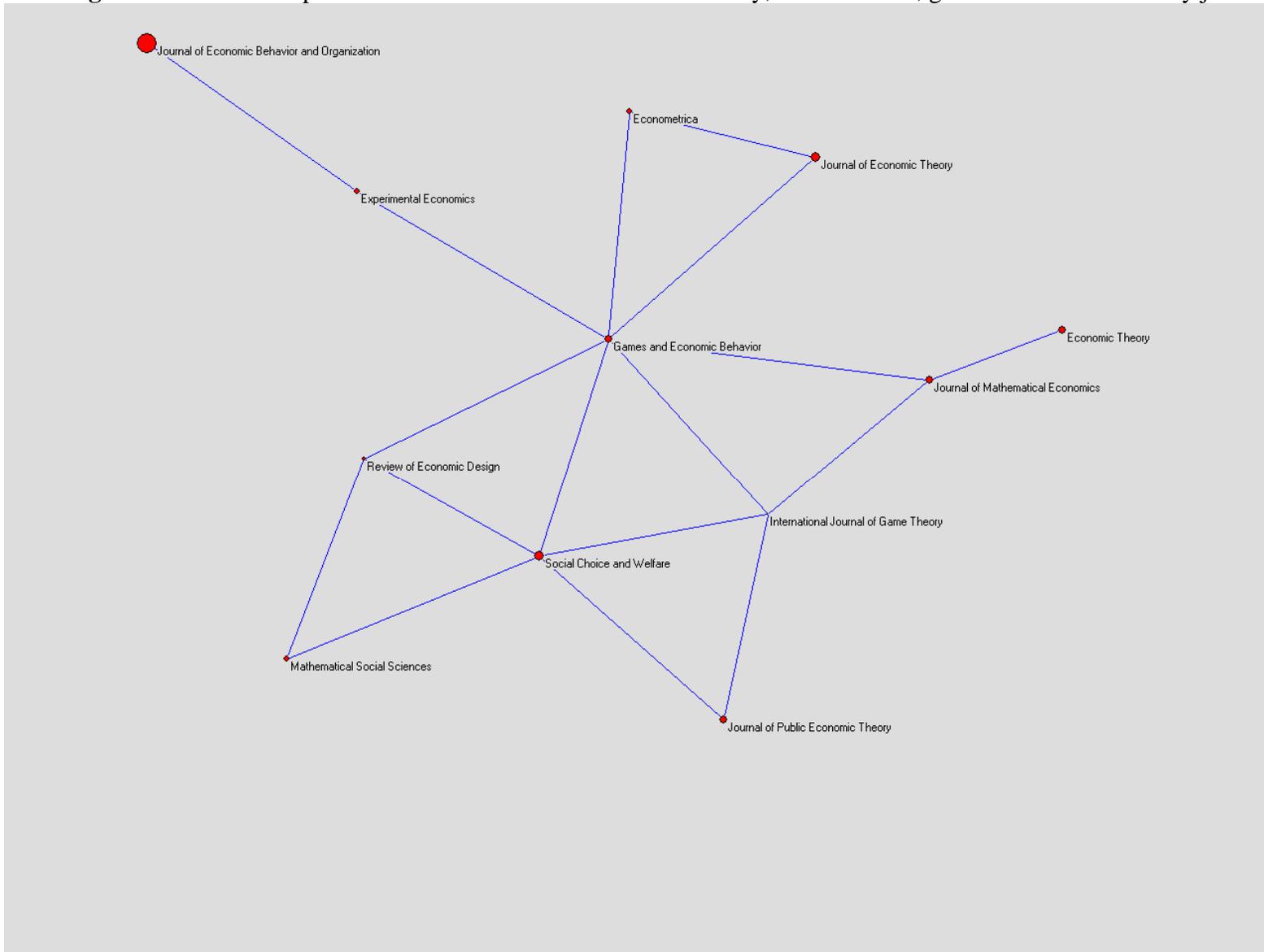

**Figure 4.** A weak component in 6-slices network: economic theory, econometrics, game and decision theory journals

**Figure 5.** Weak components in 6-slices network with more than three journals

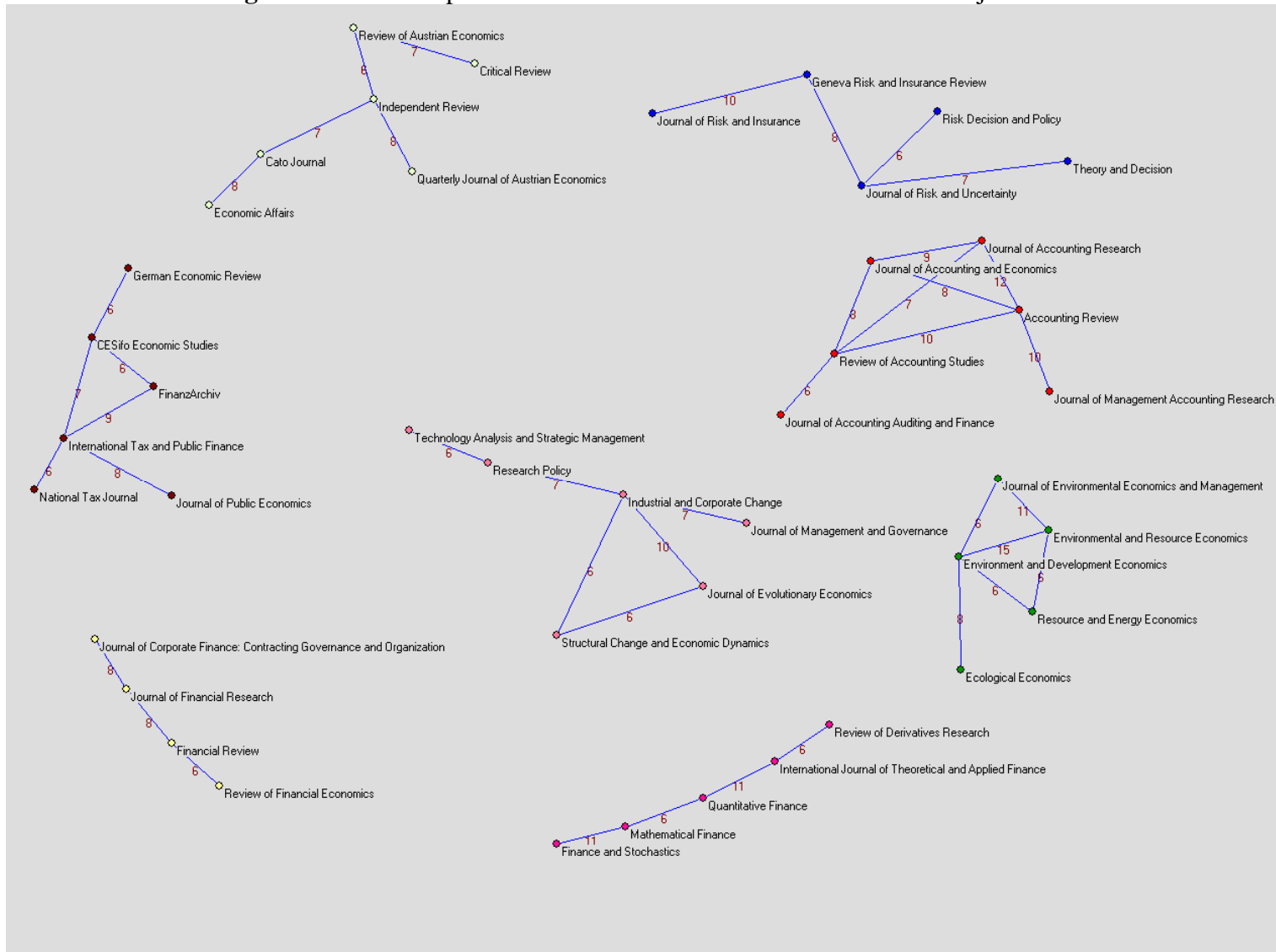

**Figure 6.** Weak components in 6-slices network with three journals

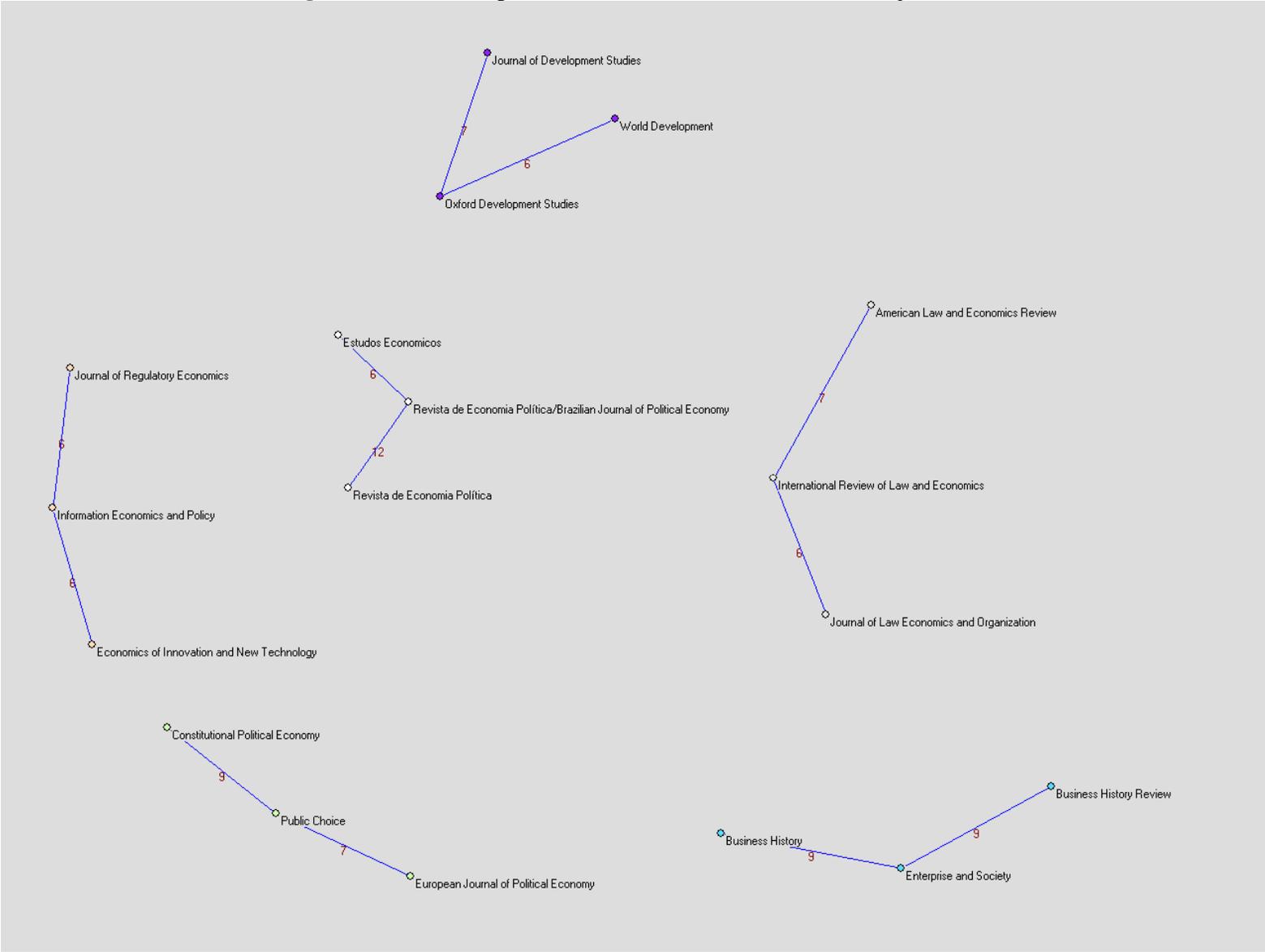

| Journal | Degree | Normalized degree | Rank degree | Centrality | Rank centrality | Betweenness (x100) | Rank betweenness |
|---|---|---|---|---|---|---|---|
| Academia Economic Papers | 8 | 0,011 | 433 | 0,323 | 334 | 0,026 | 470 |
| Accounting Business and Financial History | 6 | 0,008 | 479 | 0,303 | 431 | 0,011 | 523 |
| Accounting Review | 8 | 0,011 | 433 | 0,276 | 560 | 0,011 | 520 |
| Acta Oeconomica | 10 | 0,013 | 393 | 0,325 | 328 | 0,104 | 332 |
| African Development Review/Revue Africaine de Developpement | 26 | 0,035 | 189 | 0,351 | 193 | 0,228 | 229 |
| African Economic History | 0 | 0,000 | 673 | 0,000 | 673 | 0,000 | 593 |
| African Finance Journal | 21 | 0,028 | 235 | 0,342 | 237 | 0,307 | 169 |
| Afrika Spectrum | 3 | 0,004 | 553 | 0,282 | 531 | 0,000 | 593 |
| Agenda | 5 | 0,007 | 498 | 0,284 | 516 | 0,006 | 544 |
| Agribusiness | 8 | 0,011 | 433 | 0,282 | 531 | 0,058 | 399 |
| Agricultural and Resource Economics Review | 0 | 0,000 | 673 | 0,000 | 673 | 0,000 | 593 |
| Agricultural Economics | 11 | 0,015 | 373 | 0,293 | 472 | 0,111 | 322 |
| Agricultural Finance Review | 2 | 0,003 | 590 | 0,271 | 575 | 0,005 | 547 |
| Agriculture and Human Values | 0 | 0,000 | 673 | 0,000 | 673 | 0,000 | 593 |
| Agronomia Mesoamericana | 0 | 0,000 | 673 | 0,000 | 673 | 0,000 | 593 |
| Allgemeines Statistisches Archiv/Journal of the German Statistical Society | 2 | 0,003 | 590 | 0,258 | 607 | 0,005 | 552 |
| American Economic Review | 40 | 0,054 | 90 | 0,378 | 72 | 0,518 | 82 |
| American Economist | 12 | 0,016 | 357 | 0,328 | 311 | 0,024 | 474 |
| American Enterprise | 0 | 0,000 | 673 | 0,000 | 673 | 0,000 | 593 |
| American Historical Review | 0 | 0,000 | 673 | 0,000 | 673 | 0,000 | 593 |
| American Journal of Agricultural Economics | 15 | 0,020 | 300 | 0,321 | 342 | 0,311 | 165 |
| American Journal of Economics and Sociology | 3 | 0,004 | 553 | 0,253 | 614 | 0,000 | 593 |
| American Law and Economics Review | 19 | 0,026 | 259 | 0,341 | 243 | 0,102 | 335 |
| American Political Science Review | 7 | 0,009 | 454 | 0,285 | 513 | 0,053 | 409 |
| American Prospect | 5 | 0,007 | 498 | 0,290 | 487 | 0,005 | 550 |
| American Statistician | 1 | 0,001 | 629 | 0,003 | 671 | 0,000 | 593 |
| Analyse Prévision | 5 | 0,007 | 498 | 0,282 | 530 | 0,013 | 513 |
| Annales dEconomie et de Statistique | 34 | 0,046 | 125 | 0,361 | 149 | 0,426 | 117 |
| Annals of Economics and Finance | 44 | 0,059 | 65 | 0,371 | 98 | 0,545 | 74 |
| Annals of Public and Cooperative Economics | 13 | 0,017 | 335 | 0,327 | 314 | 0,345 | 150 |
| Annals of Regional Science | 48 | 0,064 | 47 | 0,373 | 90 | 0,984 | 23 |
| Annals of the American Academy of Political and Social Science | 3 | 0,004 | 553 | 0,280 | 543 | 0,007 | 538 |
| Antitrust Bulletin | 0 | 0,000 | 673 | 0,000 | 673 | 0,000 | 593 |
| Antitrust Law and Economics Review | 15 | 0,020 | 300 | 0,329 | 300 | 0,118 | 315 |
| Applied Economics | 36 | 0,048 | 112 | 0,371 | 100 | 0,192 | 256 |
| Applied Economics Letters | 36 | 0,048 | 112 | 0,371 | 100 | 0,192 | 256 |
| Applied Economics Quarterly | 22 | 0,030 | 224 | 0,344 | 229 | 0,229 | 227 |
| Applied Financial Economics | 29 | 0,039 | 155 | 0,361 | 148 | 0,126 | 308 |
| Applied Mathematical Finance | 5 | 0,007 | 498 | 0,280 | 543 | 0,140 | 293 |
| Aquaculture Economics and Management | 10 | 0,013 | 393 | 0,305 | 417 | 0,132 | 301 |
| Archives of Economic History | 17 | 0,023 | 280 | 0,324 | 330 | 0,094 | 345 |
| Arthaniti | 33 | 0,044 | 131 | 0,367 | 112 | 0,127 | 304 |
| ASEAN Economic Bulletin | 13 | 0,017 | 335 | 0,314 | 372 | 0,025 | 472 |
| Asia Pacific Business Review | 21 | 0,028 | 235 | 0,339 | 255 | 0,251 | 211 |
| Asia Pacific Journal of Economics and Business | 6 | 0,008 | 479 | 0,306 | 414 | 0,063 | 386 |
| Asian Development Review | 41 | 0,055 | 83 | 0,372 | 94 | 0,417 | 121 |
| Asian Economic Journal | 52 | 0,070 | 38 | 0,395 | 20 | 0,421 | 120 |
| Asian-Pacific Economic Literature | 43 | 0,058 | 72 | 0,366 | 117 | 0,437 | 111 |
| Asia-Pacific Development Journal | 7 | 0,009 | 454 | 0,300 | 451 | 0,001 | 582 |
| Asia-Pacific Financial Markets | 7 | 0,009 | 454 | 0,301 | 435 | 0,002 | 565 |
| Atlantic Economic Journal | 44 | 0,059 | 65 | 0,385 | 43 | 0,763 | 41 |
| Aussenwirtschaft | 3 | 0,004 | 553 | 0,295 | 466 | 0,002 | 567 |
| Australasian Journal of Regional Studies | 18 | 0,024 | 271 | 0,330 | 299 | 0,160 | 279 |
| Australian Bulletin of Labour | 0 | 0,000 | 673 | 0,000 | 673 | 0,000 | 593 |
| Australian Commodities Forecasts and Issues | 0 | 0,000 | 673 | 0,000 | 673 | 0,000 | 593 |
| Australian Economic History Review | 7 | 0,009 | 454 | 0,318 | 357 | 0,036 | 435 |
| Australian Economic Papers | 30 | 0,040 | 148 | 0,365 | 121 | 0,177 | 272 |
| Australian Economic Review | 25 | 0,034 | 202 | 0,358 | 161 | 0,150 | 283 |
| Australian Journal of Agricultural and Resource Economics | 12 | 0,016 | 357 | 0,328 | 310 | 0,080 | 366 |
| Australian Journal of Labour Economics | 21 | 0,028 | 235 | 0,342 | 238 | 0,224 | 232 |
| Australian Journal of Management | 3 | 0,004 | 553 | 0,286 | 505 | 0,001 | 572 |
| Banca Nazionale del Lavoro Quarterly Review | 2 | 0,003 | 590 | 0,264 | 593 | 0,000 | 593 |
| Bancaria | 0 | 0,000 | 673 | 0,000 | 673 | 0,000 | 593 |
| Bangladesh Development Studies | 35 | 0,047 | 119 | 0,370 | 106 | 0,202 | 247 |
| Bank of Israel Economic Review | 9 | 0,012 | 412 | 0,325 | 327 | 0,014 | 511 |
| Bank of Japan Monetary and Economic Studies | 14 | 0,019 | 319 | 0,332 | 281 | 0,055 | 405 |
| Bank of Valletta Review | 1 | 0,001 | 629 | 0,155 | 670 | 0,000 | 593 |
| Banker | 1 | 0,001 | 629 | 0,265 | 588 | 0,000 | 593 |
| Behavioral Research in Accounting | 6 | 0,008 | 479 | 0,288 | 497 | 0,014 | 509 |
| Brazilian Electronic Journal of Economics | 9 | 0,012 | 412 | 0,317 | 360 | 0,059 | 394 |
| Brazilian Review of Econometrics | 22 | 0,030 | 224 | 0,341 | 241 | 0,258 | 204 |
| British Journal of Industrial Relations | 14 | 0,019 | 319 | 0,330 | 292 | 0,104 | 333 |
| Brookings Papers on Economic Activity | 0 | 0,000 | 673 | 0,000 | 673 | 0,000 | 593 |
| Brookings-Wharton Papers on Financial Services | 2 | 0,003 | 590 | 0,277 | 554 | 0,001 | 578 |
| Brookings-Wharton Papers on Urban Affairs | 2 | 0,003 | 590 | 0,282 | 528 | 0,000 | 593 |
| Buffalo Law Review | 0 | 0,000 | 673 | 0,000 | 673 | 0,000 | 593 |
| Bulletin for International Fiscal Documentation | 5 | 0,007 | 498 | 0,259 | 604 | 0,247 | 215 |
| Bulletin of Economic Research | 40 | 0,054 | 90 | 0,371 | 98 | 0,492 | 89 |
| Bulletin of Indonesian Economic Studies | 22 | 0,030 | 224 | 0,332 | 284 | 0,145 | 286 |
| Business and Economic History | 2 | 0,003 | 590 | 0,252 | 619 | 0,000 | 593 |
| Business Economics | 5 | 0,007 | 498 | 0,283 | 521 | 0,003 | 563 |
| Business History | 19 | 0,026 | 259 | 0,340 | 249 | 0,481 | 93 |
| Business History Review | 7 | 0,009 | 454 | 0,308 | 404 | 0,056 | 402 |
| Cahiers dEconomie et Sociologie Rurales | 12 | 0,016 | 357 | 0,294 | 470 | 0,708 | 49 |
| Cahiers dEconomie Politique | 10 | 0,013 | 393 | 0,301 | 438 | 0,020 | 485 |
| Cahiers Economiques de Bruxelles | 1 | 0,001 | 629 | 0,258 | 605 | 0,000 | 593 |
| California Management Review | 17 | 0,023 | 280 | 0,330 | 297 | 0,120 | 312 |

| Journal | Degree | Normalized degree | Rank degree | Centrality | Rank centrality | Betweenness (x100) | Rank betweenness |
|---|---|---|---|---|---|---|---|
| Cambridge Journal of Economics | 59 | 0,079 | 22 | 0,388 | 36 | 0,974 | 24 |
| Canadian Business Economics | 7 | 0,009 | 454 | 0,324 | 331 | 0,029 | 459 |
| Canadian Journal of Agricultural Economics | 0 | 0,000 | 673 | 0,000 | 673 | 0,000 | 593 |
| Canadian Journal of Development Studies | 3 | 0,004 | 553 | 0,258 | 605 | 0,009 | 531 |
| Canadian Journal of Economics | 16 | 0,021 | 288 | 0,345 | 224 | 0,120 | 313 |
| Canadian Journal of Regional Science | 11 | 0,015 | 373 | 0,295 | 467 | 0,053 | 410 |
| Canadian Public Policy | 6 | 0,008 | 479 | 0,293 | 471 | 0,030 | 455 |
| Canadian Tax Journal | 10 | 0,013 | 393 | 0,300 | 444 | 0,082 | 364 |
| Cato Journal | 30 | 0,040 | 148 | 0,360 | 157 | 0,204 | 245 |
| Central European Journal of Operations Research | 8 | 0,011 | 433 | 0,301 | 440 | 0,071 | 377 |
| CESifo Economic Studies | 63 | 0,085 | 16 | 0,400 | 12 | 1,116 | 16 |
| Challenge | 39 | 0,052 | 96 | 0,384 | 48 | 0,368 | 140 |
| China Economic Review | 19 | 0,026 | 259 | 0,345 | 221 | 0,147 | 284 |
| China Quarterly | 1 | 0,001 | 629 | 0,222 | 656 | 0,000 | 593 |
| Chinese Economy | 3 | 0,004 | 553 | 0,280 | 540 | 0,001 | 575 |
| CIRIEC-España Revista de Economía Pública Social y Cooperativa | 3 | 0,004 | 553 | 0,283 | 521 | 0,014 | 510 |
| Comercio Exterior | 1 | 0,001 | 629 | 0,227 | 651 | 0,000 | 593 |
| Communications and Strategies | 2 | 0,003 | 590 | 0,241 | 636 | 0,000 | 593 |
| Comparative Economic Studies | 11 | 0,015 | 373 | 0,325 | 326 | 0,092 | 350 |
| Computational Economics | 27 | 0,036 | 176 | 0,363 | 132 | 0,190 | 261 |
| Conceptos (Buenos Aires) | 0 | 0,000 | 673 | 0,000 | 673 | 0,000 | 593 |
| Conflict Management and Peace Science | 8 | 0,011 | 433 | 0,309 | 394 | 0,046 | 424 |
| Constitutional Political Economy | 23 | 0,031 | 216 | 0,352 | 184 | 0,143 | 289 |
| Contemporary Economic Policy | 72 | 0,097 | 6 | 0,401 | 10 | 1,883 | 4 |
| Contributions to Political Economy | 1 | 0,001 | 629 | 0,268 | 581 | 0,000 | 593 |
| Critical Review | 39 | 0,052 | 96 | 0,376 | 76 | 0,484 | 92 |
| Cuadernos de Economía | 6 | 0,008 | 479 | 0,305 | 417 | 0,005 | 554 |
| Cuadernos Economicos de I.C.E. | 20 | 0,027 | 245 | 0,330 | 292 | 0,137 | 296 |
| Cyprus Review | 1 | 0,001 | 629 | 0,219 | 657 | 0,000 | 593 |
| De Economist | 38 | 0,051 | 102 | 0,377 | 73 | 0,290 | 182 |
| Defence and Peace Economics | 20 | 0,027 | 245 | 0,346 | 220 | 0,134 | 299 |
| Demography | 11 | 0,015 | 373 | 0,320 | 347 | 0,215 | 239 |
| Desarrollo Económico | 2 | 0,003 | 590 | 0,252 | 617 | 0,003 | 557 |
| Developing Economies | 38 | 0,051 | 102 | 0,373 | 90 | 0,347 | 148 |
| Development | 20 | 0,027 | 245 | 0,351 | 197 | 0,072 | 373 |
| Development and Change | 23 | 0,031 | 216 | 0,331 | 287 | 0,304 | 170 |
| Development Southern Africa | 6 | 0,008 | 479 | 0,299 | 456 | 0,008 | 537 |
| Eastern Africa Journal of Rural Development | 0 | 0,000 | 673 | 0,000 | 673 | 0,000 | 593 |
| Eastern Economic Journal | 20 | 0,027 | 245 | 0,335 | 273 | 0,118 | 316 |
| Eastern European Economics | 4 | 0,005 | 530 | 0,272 | 572 | 0,000 | 593 |
| East-West Journal of Economics and Business | 38 | 0,051 | 102 | 0,375 | 80 | 0,734 | 45 |
| ECLAP Review | 1 | 0,001 | 629 | 0,227 | 651 | 0,000 | 593 |
| Ecological Economics | 27 | 0,036 | 176 | 0,361 | 146 | 0,500 | 85 |
| Econometric Reviews | 19 | 0,026 | 259 | 0,339 | 258 | 0,078 | 367 |
| Econometric Theory | 9 | 0,012 | 412 | 0,308 | 399 | 0,002 | 568 |
| Econometrica | 40 | 0,054 | 90 | 0,362 | 143 | 0,239 | 222 |
| Econometrics Journal | 9 | 0,012 | 412 | 0,309 | 395 | 0,011 | 522 |
| Economia (Pontifical Catholic University of Peru) | 1 | 0,001 | 629 | 0,265 | 589 | 0,000 | 593 |
| Economia Aplicada/Brazilian Journal of Applied Economics | 22 | 0,030 | 224 | 0,333 | 280 | 0,360 | 144 |
| Economía Chilena | 12 | 0,016 | 357 | 0,320 | 346 | 0,034 | 442 |
| Economia e Lavoro | 4 | 0,005 | 530 | 0,273 | 570 | 0,126 | 307 |
| Economia Industrial | 7 | 0,009 | 454 | 0,286 | 507 | 0,037 | 433 |
| Economia Internazionale/International Economics | 1 | 0,001 | 629 | 0,219 | 658 | 0,000 | 593 |
| Economia Mexicana Nueva Epoca | 11 | 0,015 | 373 | 0,304 | 425 | 0,571 | 68 |
| Economia Politica | 15 | 0,020 | 300 | 0,332 | 281 | 0,053 | 411 |
| Economic Affairs | 36 | 0,048 | 112 | 0,363 | 136 | 0,544 | 75 |
| Economic Analysis and Policy | 2 | 0,003 | 590 | 0,259 | 602 | 0,000 | 593 |
| Economic and Business Review | 13 | 0,017 | 335 | 0,308 | 405 | 0,435 | 112 |
| Economic and Financial Modelling | 12 | 0,016 | 357 | 0,323 | 335 | 0,044 | 426 |
| Economic and Industrial Democracy | 18 | 0,024 | 271 | 0,342 | 239 | 0,290 | 181 |
| Economic and Labour Relations Review | 18 | 0,024 | 271 | 0,340 | 251 | 0,073 | 370 |
| Economic and Social Review | 10 | 0,013 | 393 | 0,316 | 365 | 0,035 | 437 |
| Economic Development and Cultural Change | 14 | 0,019 | 319 | 0,328 | 307 | 0,030 | 453 |
| Economic Development Quarterly | 14 | 0,019 | 319 | 0,301 | 440 | 0,055 | 403 |
| Economic Geography | 13 | 0,017 | 335 | 0,313 | 379 | 0,098 | 342 |
| Economic History Review | 8 | 0,011 | 433 | 0,307 | 410 | 0,009 | 532 |
| Economic Inquiry | 19 | 0,026 | 259 | 0,344 | 226 | 0,117 | 317 |
| Economic Issues | 19 | 0,026 | 259 | 0,353 | 183 | 0,104 | 330 |
| Economic Journal | 21 | 0,028 | 235 | 0,341 | 243 | 0,069 | 378 |
| Economic Modelling | 34 | 0,046 | 125 | 0,372 | 95 | 0,738 | 43 |
| Economic Notes | 49 | 0,066 | 46 | 0,378 | 70 | 0,639 | 56 |
| Economic Perspectives | 0 | 0,000 | 673 | 0,000 | 673 | 0,000 | 593 |
| Economic Policy | 38 | 0,051 | 102 | 0,368 | 111 | 0,263 | 197 |
| Economic Record | 19 | 0,026 | 259 | 0,351 | 193 | 0,110 | 323 |
| Economic Systems | 27 | 0,036 | 176 | 0,356 | 172 | 0,804 | 36 |
| Economic Systems Research | 17 | 0,023 | 280 | 0,331 | 289 | 0,280 | 190 |
| Economic Theory | 41 | 0,055 | 83 | 0,379 | 67 | 0,414 | 122 |
| Economica | 28 | 0,038 | 165 | 0,357 | 167 | 0,257 | 206 |
| Económica (National University of La Plata) | 4 | 0,005 | 530 | 0,279 | 547 | 0,029 | 458 |
| Economics and Philosophy | 18 | 0,024 | 271 | 0,341 | 246 | 0,046 | 423 |
| Economics and Politics | 29 | 0,039 | 155 | 0,376 | 74 | 0,263 | 199 |
| Economics Letters | 46 | 0,062 | 57 | 0,376 | 77 | 0,388 | 131 |
| Economics of Education Review | 15 | 0,020 | 300 | 0,327 | 316 | 0,393 | 130 |
| Economics of Governance | 26 | 0,035 | 189 | 0,351 | 193 | 0,296 | 176 |
| Economics of Innovation and New Technology | 27 | 0,036 | 176 | 0,354 | 177 | 0,204 | 246 |
| Economics of Planning (dal 9/01/06 Economic Change and Restructuring) | 28 | 0,038 | 165 | 0,357 | 167 | 0,547 | 73 |
| Economics of Transition | 11 | 0,015 | 373 | 0,318 | 358 | 0,107 | 327 |

| Journal | Degree | Normalized degree | Rank degree | Centrality | Rank centrality | Betweenness (x100) | Rank betweenness |
|---|---|---|---|---|---|---|---|
| Economie Appliquée | 14 | 0,019 | 319 | 0,318 | 356 | 0,263 | 198 |
| Economie et Prévision | 22 | 0,030 | 224 | 0,345 | 223 | 0,331 | 153 |
| Economie et Statistique | 2 | 0,003 | 590 | 0,222 | 655 | 0,000 | 593 |
| Economie Internationale | 24 | 0,032 | 208 | 0,362 | 139 | 0,477 | 94 |
| Economie Rurale | 4 | 0,005 | 530 | 0,226 | 653 | 0,252 | 210 |
| Economies et Sociétés | 7 | 0,009 | 454 | 0,290 | 487 | 0,059 | 396 |
| Economisch en Sociaal Tijdschrift | 0 | 0,000 | 673 | 0,000 | 673 | 0,000 | 593 |
| Economy and Society | 5 | 0,007 | 498 | 0,273 | 569 | 0,005 | 549 |
| Education Economics | 11 | 0,015 | 373 | 0,319 | 350 | 0,058 | 397 |
| Ekonomia | 41 | 0,055 | 83 | 0,370 | 104 | 0,368 | 139 |
| Ekonomiska Samfundets Tidskrift | 6 | 0,008 | 479 | 0,313 | 377 | 0,011 | 519 |
| Ekonomska Misao i Praksa | 0 | 0,000 | 673 | 0,000 | 673 | 0,000 | 593 |
| El Trimestre Economico | 13 | 0,017 | 335 | 0,322 | 336 | 0,232 | 225 |
| Emerging Markets Finance and Trade | 27 | 0,036 | 176 | 0,362 | 143 | 0,404 | 125 |
| Empirica | 45 | 0,060 | 62 | 0,380 | 63 | 0,465 | 97 |
| Empirical Economics | 29 | 0,039 | 155 | 0,362 | 139 | 0,409 | 124 |
| Energy Economics | 15 | 0,020 | 300 | 0,331 | 289 | 0,116 | 319 |
| Energy Journal | 14 | 0,019 | 319 | 0,326 | 318 | 0,066 | 381 |
| Energy Studies Review | 1 | 0,001 | 629 | 0,264 | 592 | 0,000 | 593 |
| Engineering Economist | 0 | 0,000 | 673 | 0,000 | 673 | 0,000 | 593 |
| Enterprise and Society | 15 | 0,020 | 300 | 0,305 | 422 | 0,218 | 237 |
| Entrepreneurship and Regional Development | 10 | 0,013 | 393 | 0,289 | 494 | 0,084 | 360 |
| Environment and Development Economics | 55 | 0,074 | 29 | 0,392 | 24 | 1,005 | 20 |
| Environment and Planning A | 13 | 0,017 | 335 | 0,312 | 382 | 0,102 | 336 |
| Environment and Planning C: Government and Policy | 13 | 0,017 | 335 | 0,315 | 370 | 0,437 | 110 |
| Environmental and Resource Economics | 47 | 0,063 | 52 | 0,385 | 43 | 0,725 | 46 |
| Environmental Economics and Policy Studies | 15 | 0,020 | 300 | 0,326 | 318 | 0,031 | 450 |
| Environmental Values | 1 | 0,001 | 629 | 0,233 | 648 | 0,000 | 593 |
| Estudios de Economia | 0 | 0,000 | 673 | 0,000 | 673 | 0,000 | 593 |
| Estudios Económicos | 15 | 0,020 | 300 | 0,333 | 278 | 0,317 | 160 |
| Estudios Internacionales | 0 | 0,000 | 673 | 0,000 | 673 | 0,000 | 593 |
| Estudos Economicos | 6 | 0,008 | 479 | 0,285 | 515 | 0,018 | 495 |
| Eurasian Geography and Economics | 7 | 0,009 | 454 | 0,276 | 564 | 0,122 | 311 |
| EuroChoices | 5 | 0,007 | 498 | 0,283 | 525 | 0,055 | 404 |
| European Economic Review | 50 | 0,067 | 43 | 0,383 | 54 | 0,775 | 38 |
| European Financial Management | 60 | 0,081 | 20 | 0,390 | 30 | 0,936 | 26 |
| European Journal of Development Research | 19 | 0,026 | 259 | 0,329 | 302 | 0,186 | 265 |
| European Journal of Finance | 47 | 0,063 | 52 | 0,381 | 58 | 1,001 | 21 |
| European Journal of Housing Policy | 12 | 0,016 | 357 | 0,325 | 323 | 0,009 | 529 |
| European Journal of Industrial Relations | 8 | 0,011 | 433 | 0,284 | 517 | 0,030 | 456 |
| European Journal of International Relations | 16 | 0,021 | 288 | 0,311 | 386 | 0,335 | 152 |
| European Journal of Law and Economics | 27 | 0,036 | 176 | 0,357 | 166 | 0,737 | 44 |
| European Journal of Political Economy | 69 | 0,093 | 10 | 0,402 | 9 | 1,362 | 10 |
| European Journal of the History of Economic Thought | 43 | 0,058 | 72 | 0,379 | 67 | 0,859 | 30 |
| European Review of Agricultural Economics | 5 | 0,007 | 498 | 0,277 | 556 | 0,017 | 499 |
| European Review of Economic History | 17 | 0,023 | 280 | 0,316 | 364 | 0,345 | 149 |
| Experimental Economics | 41 | 0,055 | 83 | 0,385 | 42 | 0,285 | 186 |
| Explorations in Economic History | 15 | 0,020 | 300 | 0,337 | 263 | 0,207 | 244 |
| Family Economics and Nutrition Review | 1 | 0,001 | 629 | 0,214 | 660 | 0,000 | 593 |
| FDIC Banking Review | 0 | 0,000 | 673 | 0,000 | 673 | 0,000 | 593 |
| Federal Reserve Bank of Atlanta Economic Review | 4 | 0,005 | 530 | 0,284 | 518 | 0,001 | 577 |
| Federal Reserve Bank of Chicago Economic Perspectives | 4 | 0,005 | 530 | 0,299 | 456 | 0,008 | 534 |
| Federal Reserve Bank of Dallas Economic and Financial Policy Review | 5 | 0,007 | 498 | 0,282 | 527 | 0,000 | 593 |
| Federal Reserve Bank of Kansas City Economic Review | 4 | 0,005 | 530 | 0,301 | 438 | 0,000 | 588 |
| Federal Reserve Bank of Minneapolis Quarterly Review | 3 | 0,004 | 553 | 0,297 | 460 | 0,000 | 593 |
| Federal Reserve Bank of New York Economic Policy Review | 44 | 0,059 | 65 | 0,366 | 117 | 0,439 | 109 |
| Federal Reserve Bank of San Francisco Economic Review | 2 | 0,003 | 590 | 0,287 | 498 | 0,000 | 592 |
| Federal Reserve Bank of St. Louis Review | 8 | 0,011 | 433 | 0,311 | 383 | 0,003 | 561 |
| Federal Reserve Bulletin | 4 | 0,005 | 530 | 0,286 | 509 | 0,000 | 593 |
| Feminist Economics | 54 | 0,072 | 31 | 0,385 | 45 | 1,192 | 13 |
| Finance | 27 | 0,036 | 176 | 0,339 | 257 | 0,072 | 375 |
| Finance a Úver/Czech Journal of Economics and Finance | 7 | 0,009 | 454 | 0,277 | 554 | 0,026 | 469 |
| Finance and Development | 5 | 0,007 | 498 | 0,284 | 518 | 0,074 | 369 |
| Finance and Stochastics | 20 | 0,027 | 245 | 0,326 | 318 | 0,073 | 371 |
| Finance India | 72 | 0,097 | 6 | 0,405 | 5 | 1,449 | 9 |
| Financial History Review | 4 | 0,005 | 530 | 0,266 | 584 | 0,003 | 560 |
| Financial Management | 6 | 0,008 | 479 | 0,290 | 491 | 0,000 | 593 |
| Financial Markets Institutions and Instruments | 5 | 0,007 | 498 | 0,300 | 451 | 0,000 | 593 |
| Financial Review | 40 | 0,054 | 90 | 0,362 | 139 | 0,177 | 271 |
| FinanzArchiv | 38 | 0,051 | 102 | 0,371 | 100 | 0,396 | 128 |
| Finnish Economic Papers | 21 | 0,028 | 235 | 0,349 | 204 | 0,170 | 278 |
| Fiscal Studies | 33 | 0,044 | 131 | 0,358 | 162 | 0,232 | 226 |
| Food Policy | 6 | 0,008 | 479 | 0,299 | 455 | 0,023 | 479 |
| Foreign Affairs | 3 | 0,004 | 553 | 0,259 | 600 | 0,019 | 493 |
| Foresight | 9 | 0,012 | 412 | 0,293 | 475 | 0,064 | 383 |
| Forum for Development Studies | 0 | 0,000 | 673 | 0,000 | 673 | 0,000 | 593 |
| Forum for Social Economics | 2 | 0,003 | 590 | 0,264 | 591 | 0,002 | 566 |
| Games and Economic Behavior | 47 | 0,063 | 52 | 0,384 | 48 | 0,461 | 99 |
| Geneva Papers on Risk and Insurance: Issues and Practice | 7 | 0,009 | 454 | 0,306 | 412 | 0,002 | 570 |
| Geneva Risk and Insurance Review | 39 | 0,052 | 96 | 0,376 | 79 | 0,329 | 155 |
| German Economic Review | 65 | 0,087 | 14 | 0,399 | 16 | 0,884 | 29 |
| Gestion 2000 | 2 | 0,003 | 590 | 0,248 | 624 | 0,003 | 564 |
| Giornale degli Economisti e Annali di Economia | 11 | 0,015 | 373 | 0,307 | 406 | 0,028 | 462 |
| Global Business and Economics Review | 15 | 0,020 | 300 | 0,332 | 281 | 0,175 | 274 |
| Global Economic Review | 23 | 0,031 | 216 | 0,359 | 160 | 0,045 | 425 |
| Global Economy Quarterly | 17 | 0,023 | 280 | 0,338 | 262 | 0,252 | 209 |
| Global Environmental Politics | 12 | 0,016 | 357 | 0,296 | 462 | 0,099 | 340 |

| Journal | Degree | Normalized degree | Rank degree | Centrality | Rank centrality | Betweenness (x100) | Rank betweenness |
|---|---|---|---|---|---|---|---|
| Global Finance Journal | 23 | 0,031 | 216 | 0,338 | 261 | 0,027 | 465 |
| Growth and Change | 18 | 0,024 | 271 | 0,314 | 375 | 0,093 | 349 |
| Harvard Business Review | 0 | 0,000 | 673 | 0,000 | 673 | 0,000 | 593 |
| Health Care Management Science | 4 | 0,005 | 530 | 0,281 | 537 | 0,017 | 500 |
| Health Economics | 16 | 0,021 | 288 | 0,326 | 321 | 0,220 | 236 |
| Health Marketing Quarterly | 5 | 0,007 | 498 | 0,293 | 475 | 0,031 | 452 |
| Health Services Research | 9 | 0,012 | 412 | 0,310 | 389 | 0,025 | 473 |
| History of Economic Ideas | 30 | 0,040 | 148 | 0,347 | 214 | 0,380 | 135 |
| History of Economics Review | 14 | 0,019 | 319 | 0,316 | 367 | 0,016 | 502 |
| History of Political Economy | 12 | 0,016 | 357 | 0,311 | 386 | 0,057 | 400 |
| Hitotsubashi Journal of Economics | 2 | 0,003 | 590 | 0,292 | 479 | 0,000 | 587 |
| Housing Policy Debate | 21 | 0,028 | 235 | 0,339 | 254 | 0,434 | 114 |
| Housing Studies | 11 | 0,015 | 373 | 0,291 | 483 | 0,080 | 365 |
| Human Resource Development Quarterly | 0 | 0,000 | 673 | 0,000 | 673 | 0,000 | 593 |
| Humanomics | 2 | 0,003 | 590 | 0,275 | 566 | 0,000 | 593 |
| IIUM Journal of Economics and Management | 3 | 0,004 | 553 | 0,277 | 552 | 0,000 | 593 |
| Il Pensiero Economico Italiano | 7 | 0,009 | 454 | 0,279 | 545 | 0,020 | 484 |
| Il Politico | 1 | 0,001 | 629 | 0,248 | 623 | 0,000 | 593 |
| Il Risparmio | 3 | 0,004 | 553 | 0,252 | 618 | 0,068 | 380 |
| IMF Staff Papers | 5 | 0,007 | 498 | 0,292 | 480 | 0,010 | 527 |
| Independent Review | 29 | 0,039 | 155 | 0,352 | 187 | 0,299 | 174 |
| Indian Economic and Social History Review | 20 | 0,027 | 245 | 0,347 | 212 | 0,228 | 228 |
| Indian Economic Journal | 28 | 0,038 | 165 | 0,369 | 109 | 0,362 | 143 |
| Indian Economic Review | 23 | 0,031 | 216 | 0,344 | 230 | 0,128 | 303 |
| Indian Journal of Economics | 2 | 0,003 | 590 | 0,265 | 589 | 0,000 | 593 |
| Indian Journal of Gender Studies | 5 | 0,007 | 498 | 0,305 | 424 | 0,006 | 545 |
| Indian Journal of Labour Economics | 10 | 0,013 | 393 | 0,305 | 422 | 0,041 | 431 |
| Indiana Business Review | 0 | 0,000 | 673 | 0,000 | 673 | 0,000 | 593 |
| Industrial and Corporate Change | 61 | 0,082 | 19 | 0,401 | 10 | 1,329 | 11 |
| Industrial and Labor Relations Review | 15 | 0,020 | 300 | 0,319 | 350 | 0,094 | 347 |
| Industrial Development. Global Report/Unido | 12 | 0,016 | 357 | 0,320 | 347 | 0,020 | 492 |
| Industrial Relations | 14 | 0,019 | 319 | 0,322 | 336 | 0,091 | 351 |
| Industry and Innovation | 24 | 0,032 | 208 | 0,347 | 216 | 0,286 | 184 |
| Info | 8 | 0,011 | 433 | 0,286 | 508 | 0,100 | 339 |
| Informacion Comercial Española Revista de Economia | 6 | 0,008 | 479 | 0,286 | 511 | 0,032 | 448 |
| Information Economics and Policy | 30 | 0,040 | 148 | 0,354 | 178 | 0,309 | 168 |
| Innovations | 33 | 0,044 | 131 | 0,348 | 205 | 0,430 | 116 |
| Inquiry | 0 | 0,000 | 673 | 0,000 | 673 | 0,000 | 593 |
| Insurance: Mathematics and Economics | 11 | 0,015 | 373 | 0,308 | 401 | 0,084 | 361 |
| Integration and Trade | 7 | 0,009 | 454 | 0,289 | 492 | 0,072 | 372 |
| International Advances in Economic Research | 8 | 0,011 | 433 | 0,306 | 415 | 0,018 | 496 |
| International Economic Journal | 62 | 0,083 | 17 | 0,400 | 14 | 0,896 | 27 |
| International Economic Review | 7 | 0,009 | 454 | 0,310 | 393 | 0,035 | 438 |
| International Economy | 20 | 0,027 | 245 | 0,344 | 226 | 0,281 | 189 |
| International Finance | 48 | 0,064 | 47 | 0,389 | 33 | 0,319 | 159 |
| International Game Theory Review | 26 | 0,035 | 189 | 0,359 | 159 | 0,286 | 185 |
| International Journal of Business | 36 | 0,048 | 112 | 0,366 | 120 | 0,584 | 65 |
| International Journal of Finance and Economics | 62 | 0,083 | 17 | 0,390 | 30 | 0,773 | 40 |
| International Journal of Forecasting | 30 | 0,040 | 148 | 0,360 | 151 | 0,178 | 270 |
| International Journal of Game Theory | 20 | 0,027 | 245 | 0,346 | 218 | 0,017 | 498 |
| International Journal of Industrial Organization | 41 | 0,055 | 83 | 0,375 | 85 | 0,463 | 98 |
| International Journal of Manpower | 12 | 0,016 | 357 | 0,312 | 380 | 0,072 | 374 |
| International Journal of Production Economics | 7 | 0,009 | 454 | 0,308 | 401 | 0,159 | 280 |
| International Journal of Social Economics | 11 | 0,015 | 373 | 0,322 | 336 | 0,314 | 164 |
| International Journal of the Economics of Business | 34 | 0,046 | 125 | 0,370 | 104 | 0,497 | 88 |
| International Journal of Theoretical and Applied Finance | 32 | 0,043 | 136 | 0,356 | 170 | 0,259 | 201 |
| International Journal of Transport Economics | 9 | 0,012 | 412 | 0,290 | 490 | 0,283 | 187 |
| International Journal of Urban and Regional Research | 3 | 0,004 | 553 | 0,264 | 593 | 0,010 | 528 |
| International Labour Review | 0 | 0,000 | 673 | 0,000 | 673 | 0,000 | 593 |
| International Organization | 10 | 0,013 | 393 | 0,294 | 468 | 0,275 | 191 |
| International Regional Science Review | 31 | 0,042 | 145 | 0,341 | 246 | 0,616 | 59 |
| International Review of Applied Economics | 35 | 0,047 | 119 | 0,360 | 151 | 0,255 | 208 |
| International Review of Economics and Finance | 57 | 0,077 | 24 | 0,391 | 26 | 0,440 | 107 |
| International Review of Financial Analysis | 41 | 0,055 | 83 | 0,356 | 175 | 0,315 | 161 |
| International Review of Law and Economics | 15 | 0,020 | 300 | 0,331 | 289 | 0,093 | 348 |
| International Social Science Journal | 2 | 0,003 | 590 | 0,268 | 582 | 0,001 | 583 |
| International Tax and Public Finance | 38 | 0,051 | 102 | 0,367 | 115 | 0,225 | 231 |
| International Trade Journal | 12 | 0,016 | 357 | 0,330 | 295 | 0,035 | 439 |
| Investigación Económica | 0 | 0,000 | 673 | 0,000 | 673 | 0,000 | 593 |
| Investigaciones Economicas | 8 | 0,011 | 433 | 0,309 | 396 | 0,016 | 504 |
| Investment Policy | 0 | 0,000 | 673 | 0,000 | 673 | 0,000 | 593 |
| Irish Journal of Agricultural and Food Research | 4 | 0,005 | 530 | 0,280 | 542 | 0,005 | 548 |
| ISE Review | 11 | 0,015 | 373 | 0,297 | 458 | 0,020 | 488 |
| Jahrbuch für Regionalwissenschaft/Review of Regional Research | 3 | 0,004 | 553 | 0,272 | 572 | 0,001 | 581 |
| Jahrbücher für Nationalökonomie und Statistik | 4 | 0,005 | 530 | 0,278 | 550 | 0,010 | 524 |
| Japan and the World Economy | 54 | 0,072 | 31 | 0,394 | 22 | 0,566 | 70 |
| Japanese Economic Review | 59 | 0,079 | 22 | 0,400 | 14 | 0,896 | 28 |
| Japanese Economy | 4 | 0,005 | 530 | 0,287 | 502 | 0,000 | 593 |
| Journal for Studies in Economics and Econometrics | 1 | 0,001 | 629 | 0,248 | 625 | 0,000 | 593 |
| Journal of Accounting and Economics | 15 | 0,020 | 300 | 0,320 | 344 | 0,126 | 309 |
| Journal of Accounting Auditing and Finance | 10 | 0,013 | 393 | 0,308 | 401 | 0,049 | 418 |
| Journal of Accounting Research | 11 | 0,015 | 373 | 0,309 | 397 | 0,052 | 414 |
| Journal of African Business | 14 | 0,019 | 319 | 0,322 | 336 | 0,031 | 451 |
| Journal of African Economies | 26 | 0,035 | 189 | 0,355 | 176 | 0,249 | 213 |
| Journal of African Finance and Economic Development | 14 | 0,019 | 319 | 0,324 | 333 | 0,038 | 432 |
| Journal of Agricultural and Applied Economics | 5 | 0,007 | 498 | 0,252 | 615 | 0,009 | 530 |
| Journal of Agricultural and Resource Economics | 9 | 0,012 | 412 | 0,300 | 447 | 0,047 | 422 |

| Journal | Degree | Normalized degree | Rank degree | Centrality | Rank centrality | Betweenness (x100) | Rank betweenness |
|---|---|---|---|---|---|---|---|
| Journal of Agricultural Economics | 10 | 0,013 | 393 | 0,297 | 459 | 0,202 | 248 |
| Journal of Applied Business Research | 9 | 0,012 | 412 | 0,319 | 350 | 0,042 | 430 |
| Journal of Applied Econometrics | 57 | 0,077 | 24 | 0,384 | 47 | 0,813 | 34 |
| Journal of Applied Economics | 20 | 0,027 | 245 | 0,353 | 179 | 0,064 | 382 |
| Journal of Applied Finance | 29 | 0,039 | 155 | 0,341 | 246 | 0,095 | 344 |
| Journal of Applied Statistics | 1 | 0,001 | 629 | 0,237 | 639 | 0,000 | 593 |
| Journal of Asian Economics | 46 | 0,062 | 57 | 0,382 | 55 | 0,850 | 31 |
| Journal of Asian Studies | 0 | 0,000 | 673 | 0,000 | 673 | 0,000 | 593 |
| Journal of Asia-Pacific Business | 19 | 0,026 | 259 | 0,328 | 307 | 0,064 | 385 |
| Journal of Australian Political Economy | 3 | 0,004 | 553 | 0,286 | 509 | 0,003 | 559 |
| Journal of Banking and Finance | 72 | 0,097 | 6 | 0,392 | 25 | 1,137 | 15 |
| Journal of Bioeconomics | 30 | 0,040 | 148 | 0,367 | 114 | 0,431 | 115 |
| Journal of Business | 6 | 0,008 | 479 | 0,294 | 469 | 0,026 | 468 |
| Journal of Business and Economic Statistics | 26 | 0,035 | 189 | 0,350 | 201 | 0,251 | 212 |
| Journal of Business Research | 27 | 0,036 | 176 | 0,351 | 192 | 0,498 | 87 |
| Journal of Common Market Studies | 15 | 0,020 | 300 | 0,329 | 302 | 0,188 | 263 |
| Journal of Comparative Economics | 8 | 0,011 | 433 | 0,328 | 309 | 0,053 | 412 |
| Journal of Conflict Resolution | 22 | 0,030 | 224 | 0,344 | 230 | 0,259 | 200 |
| Journal of Consumer Affairs | 6 | 0,008 | 479 | 0,265 | 585 | 0,042 | 429 |
| Journal of Consumer Policy | 5 | 0,007 | 498 | 0,288 | 496 | 0,035 | 436 |
| Journal of Consumer Research | 10 | 0,013 | 393 | 0,300 | 447 | 0,094 | 346 |
| Journal of Corporate Finance: Contracting Governance and Organization | 37 | 0,050 | 110 | 0,348 | 209 | 0,288 | 183 |
| Journal of Cultural Economics | 26 | 0,035 | 189 | 0,360 | 156 | 0,213 | 240 |
| Journal of Derivatives | 3 | 0,004 | 553 | 0,276 | 560 | 0,000 | 593 |
| Journal of Developing Areas | 28 | 0,038 | 165 | 0,353 | 181 | 0,190 | 260 |
| Journal of Development and Economic Policies | 1 | 0,001 | 629 | 0,247 | 626 | 0,000 | 593 |
| Journal of Development Economics | 35 | 0,047 | 119 | 0,363 | 130 | 0,142 | 290 |
| Journal of Development Studies | 26 | 0,035 | 189 | 0,339 | 258 | 0,182 | 266 |
| Journal of East-West Business | 13 | 0,017 | 335 | 0,309 | 398 | 0,290 | 180 |
| Journal of Econometrics | 36 | 0,048 | 112 | 0,363 | 132 | 0,323 | 158 |
| Journal of Economic and Social Measurement | 16 | 0,021 | 288 | 0,331 | 287 | 0,037 | 434 |
| Journal of Economic and Social Policy | 15 | 0,020 | 300 | 0,325 | 323 | 0,029 | 460 |
| Journal of Economic and Social Research | 3 | 0,004 | 553 | 0,285 | 514 | 0,002 | 571 |
| Journal of Economic Behavior and Organization | 79 | 0,106 | 3 | 0,417 | 3 | 2,277 | 3 |
| Journal of Economic Development | 26 | 0,035 | 189 | 0,362 | 137 | 0,117 | 318 |
| Journal of Economic Dynamics and Control | 51 | 0,068 | 39 | 0,393 | 23 | 0,805 | 35 |
| Journal of Economic Education | 28 | 0,038 | 165 | 0,364 | 126 | 0,245 | 218 |
| Journal of Economic Growth | 51 | 0,068 | 39 | 0,386 | 40 | 0,584 | 66 |
| Journal of Economic History | 7 | 0,009 | 454 | 0,296 | 464 | 0,033 | 445 |
| Journal of Economic Integration | 38 | 0,051 | 102 | 0,370 | 103 | 0,196 | 255 |
| Journal of Economic Issues | 3 | 0,004 | 553 | 0,283 | 521 | 0,000 | 593 |
| Journal of Economic Literature | 53 | 0,071 | 35 | 0,389 | 34 | 0,544 | 76 |
| Journal of Economic Methodology | 45 | 0,060 | 62 | 0,375 | 82 | 0,597 | 63 |
| Journal of Economic Perspectives | 29 | 0,039 | 155 | 0,360 | 154 | 0,098 | 343 |
| Journal of Economic Psychology | 26 | 0,035 | 189 | 0,360 | 151 | 0,301 | 172 |
| Journal of Economic Studies | 43 | 0,058 | 72 | 0,383 | 51 | 0,700 | 50 |
| Journal of Economic Surveys | 32 | 0,043 | 136 | 0,371 | 96 | 0,446 | 104 |
| Journal of Economic Theory | 57 | 0,077 | 24 | 0,388 | 37 | 0,650 | 54 |
| Journal of Economics (MVEA) | 0 | 0,000 | 673 | 0,000 | 673 | 0,000 | 593 |
| Journal of Economics (Zeitschrift für Nationalökonomie) | 24 | 0,032 | 208 | 0,352 | 187 | 0,090 | 353 |
| Journal of Economics and Business | 37 | 0,050 | 110 | 0,360 | 154 | 0,435 | 113 |
| Journal of Economics and Finance | 24 | 0,032 | 208 | 0,347 | 215 | 0,170 | 277 |
| Journal of Economics and Management Strategy | 26 | 0,035 | 189 | 0,353 | 181 | 0,384 | 132 |
| Journal of Education Finance | 0 | 0,000 | 673 | 0,000 | 673 | 0,000 | 593 |
| Journal of Emerging Markets | 14 | 0,019 | 319 | 0,317 | 360 | 0,059 | 393 |
| Journal of Empirical Finance | 51 | 0,068 | 39 | 0,381 | 61 | 0,526 | 79 |
| Journal of Energy and Development | 4 | 0,005 | 530 | 0,288 | 495 | 0,018 | 494 |
| Journal of Energy Literature | 6 | 0,008 | 479 | 0,281 | 539 | 0,023 | 475 |
| Journal of Environment and Development | 13 | 0,017 | 335 | 0,318 | 358 | 0,098 | 341 |
| Journal of Environmental Economics and Management | 28 | 0,038 | 165 | 0,367 | 112 | 0,469 | 95 |
| Journal of Environmental Planning and Management | 10 | 0,013 | 393 | 0,305 | 417 | 0,244 | 219 |
| Journal of European Economic History | 7 | 0,009 | 454 | 0,286 | 505 | 0,027 | 466 |
| Journal of Evolutionary Economics | 47 | 0,063 | 52 | 0,375 | 80 | 0,723 | 47 |
| Journal of Family and Economic Issues | 4 | 0,005 | 530 | 0,281 | 534 | 0,257 | 207 |
| Journal of Finance | 27 | 0,036 | 176 | 0,351 | 193 | 0,061 | 390 |
| Journal of Financial and Quantitative Analysis | 32 | 0,043 | 136 | 0,348 | 205 | 0,142 | 291 |
| Journal of Financial Economics | 40 | 0,054 | 90 | 0,356 | 174 | 0,268 | 195 |
| Journal of Financial Intermediation | 25 | 0,034 | 202 | 0,335 | 271 | 0,108 | 324 |
| Journal of Financial Management and Analysis | 1 | 0,001 | 629 | 0,260 | 598 | 0,000 | 593 |
| Journal of Financial Research | 34 | 0,046 | 125 | 0,347 | 211 | 0,105 | 329 |
| Journal of Financial Services Research | 54 | 0,072 | 31 | 0,373 | 93 | 0,443 | 105 |
| Journal of Forensic Economics | 4 | 0,005 | 530 | 0,290 | 484 | 0,005 | 551 |
| Journal of Futures Markets | 3 | 0,004 | 553 | 0,280 | 540 | 0,000 | 590 |
| Journal of Geographical Systems | 20 | 0,027 | 245 | 0,337 | 263 | 0,189 | 262 |
| Journal of Health Economics | 16 | 0,021 | 288 | 0,325 | 325 | 0,157 | 282 |
| Journal of Health Politics Policy and Law | 11 | 0,015 | 373 | 0,329 | 301 | 0,101 | 337 |
| Journal of Higher Education Policy and Management | 1 | 0,001 | 629 | 0,243 | 631 | 0,000 | 593 |
| Journal of Housing Economics | 35 | 0,047 | 119 | 0,369 | 110 | 0,574 | 67 |
| Journal of Housing Research | 27 | 0,036 | 176 | 0,352 | 186 | 0,212 | 241 |
| Journal of Human Resources | 13 | 0,017 | 335 | 0,319 | 354 | 0,034 | 444 |
| Journal of Income Distribution | 9 | 0,012 | 412 | 0,305 | 421 | 0,023 | 477 |
| Journal of Industrial Economics | 26 | 0,035 | 189 | 0,349 | 202 | 0,297 | 175 |
| Journal of Institutional and Theoretical Economics | 10 | 0,013 | 393 | 0,322 | 341 | 0,016 | 505 |
| Journal of International Business Studies | 71 | 0,095 | 9 | 0,406 | 4 | 2,583 | 2 |
| Journal of International Development | 16 | 0,021 | 288 | 0,331 | 286 | 0,314 | 163 |
| Journal of International Economic Law | 29 | 0,039 | 155 | 0,356 | 172 | 0,792 | 37 |
| Journal of International Economics | 46 | 0,062 | 57 | 0,384 | 48 | 0,526 | 80 |

| Journal | Degree | Normalized degree | Rank degree | Centrality | Rank centrality | Betweenness (x100) | Rank betweenness |
|---|---|---|---|---|---|---|---|
| Journal of International Financial Markets Institutions and Money | 51 | 0,068 | 39 | 0,380 | 62 | 0,366 | 142 |
| Journal of International Money and Finance | 35 | 0,047 | 119 | 0,371 | 97 | 0,120 | 314 |
| Journal of International Trade and Economic Development | 60 | 0,081 | 20 | 0,398 | 17 | 0,466 | 96 |
| Journal of Labor Economics | 20 | 0,027 | 245 | 0,344 | 226 | 0,090 | 354 |
| Journal of Labor Research | 9 | 0,012 | 412 | 0,299 | 454 | 0,023 | 476 |
| Journal of Law and Economics | 8 | 0,011 | 433 | 0,313 | 376 | 0,059 | 395 |
| Journal of Law Economics and Organization | 22 | 0,030 | 224 | 0,340 | 249 | 0,103 | 334 |
| Journal of Legal Economics | 0 | 0,000 | 673 | 0,000 | 673 | 0,000 | 593 |
| Journal of Legal Studies | 2 | 0,003 | 590 | 0,260 | 596 | 0,001 | 573 |
| Journal of Macroeconomics | 44 | 0,059 | 65 | 0,379 | 67 | 0,381 | 134 |
| Journal of Management Accounting Research | 9 | 0,012 | 412 | 0,287 | 498 | 0,029 | 457 |
| Journal of Management and Governance | 18 | 0,024 | 271 | 0,348 | 210 | 0,191 | 258 |
| Journal of Marketing | 1 | 0,001 | 629 | 0,225 | 654 | 0,000 | 593 |
| Journal of Marketing Research | 5 | 0,007 | 498 | 0,270 | 577 | 0,003 | 558 |
| Journal of Markets and Morality | 9 | 0,012 | 412 | 0,296 | 463 | 0,005 | 553 |
| Journal of Mathematical Economics | 39 | 0,052 | 96 | 0,376 | 74 | 0,304 | 171 |
| Journal of Monetary Economics | 28 | 0,038 | 165 | 0,357 | 164 | 0,053 | 407 |
| Journal of Money Credit and Banking | 76 | 0,102 | 4 | 0,398 | 18 | 1,714 | 5 |
| Journal of Multinational Financial Management | 29 | 0,039 | 155 | 0,346 | 217 | 0,047 | 421 |
| Journal of Peace Research | 18 | 0,024 | 271 | 0,327 | 316 | 0,127 | 306 |
| Journal of Pharmaceutical Finance  Economics and Policy | 7 | 0,009 | 454 | 0,314 | 374 | 0,007 | 540 |
| Journal of Policy Analysis and Management | 11 | 0,015 | 373 | 0,304 | 428 | 0,106 | 328 |
| Journal of Policy Modeling | 48 | 0,064 | 47 | 0,383 | 51 | 0,440 | 108 |
| Journal of Policy Reform | 43 | 0,058 | 72 | 0,375 | 82 | 0,412 | 123 |
| Journal of Political Economy | 3 | 0,004 | 553 | 0,279 | 546 | 0,001 | 576 |
| Journal of Population Economics | 47 | 0,063 | 52 | 0,386 | 41 | 0,971 | 25 |
| Journal of Portfolio Management | 33 | 0,044 | 131 | 0,361 | 146 | 0,264 | 196 |
| Journal of Post Keynesian Economics | 36 | 0,048 | 112 | 0,365 | 124 | 0,621 | 58 |
| Journal of Private Enterprise | 5 | 0,007 | 498 | 0,287 | 504 | 0,000 | 591 |
| Journal of Productivity Analysis | 25 | 0,034 | 202 | 0,366 | 119 | 0,448 | 102 |
| Journal of Public and International Affairs | 0 | 0,000 | 673 | 0,000 | 673 | 0,000 | 593 |
| Journal of Public Economic Theory | 50 | 0,067 | 43 | 0,389 | 34 | 0,454 | 100 |
| Journal of Public Economics | 46 | 0,062 | 57 | 0,373 | 92 | 0,452 | 101 |
| Journal of Public Finance and Public Choice/Economia Delle Scelte Pubbliche | 5 | 0,007 | 498 | 0,289 | 492 | 0,021 | 482 |
| Journal of Real Estate Finance and Economics | 33 | 0,044 | 131 | 0,364 | 129 | 0,393 | 129 |
| Journal of Real Estate Literature | 11 | 0,015 | 373 | 0,306 | 412 | 0,004 | 555 |
| Journal of Real Estate Portfolio Management | 10 | 0,013 | 393 | 0,305 | 416 | 0,016 | 506 |
| Journal of Real Estate Practice and Education | 8 | 0,011 | 433 | 0,297 | 460 | 0,006 | 543 |
| Journal of Real Estate Research | 21 | 0,028 | 235 | 0,329 | 302 | 0,089 | 355 |
| Journal of Regional Analysis and Policy | 5 | 0,007 | 498 | 0,265 | 586 | 0,051 | 417 |
| Journal of Regional Science | 53 | 0,071 | 35 | 0,385 | 45 | 1,268 | 12 |
| Journal of Regulatory Economics | 38 | 0,051 | 102 | 0,374 | 86 | 0,651 | 53 |
| Journal of Risk and Insurance | 32 | 0,043 | 136 | 0,361 | 150 | 0,369 | 138 |
| Journal of Risk and Uncertainty | 53 | 0,071 | 35 | 0,390 | 28 | 0,585 | 64 |
| Journal of Social and Economic Development | 8 | 0,011 | 433 | 0,307 | 409 | 0,182 | 267 |
| Journal of Socio-Economics | 39 | 0,052 | 96 | 0,381 | 58 | 0,614 | 60 |
| Journal of Sports Economics | 22 | 0,030 | 224 | 0,348 | 205 | 0,133 | 300 |
| Journal of Taxation | 1 | 0,001 | 629 | 0,201 | 666 | 0,000 | 593 |
| Journal of Technology Transfer | 12 | 0,016 | 357 | 0,333 | 278 | 0,061 | 391 |
| Journal of the American Statistical Association | 1 | 0,001 | 629 | 0,003 | 671 | 0,000 | 593 |
| Journal of the Asia Pacific Economy | 46 | 0,062 | 57 | 0,388 | 37 | 0,599 | 62 |
| Journal of the History of Economic Thought | 16 | 0,021 | 288 | 0,329 | 305 | 0,114 | 320 |
| Journal of the Japanese and International Economies | 42 | 0,056 | 78 | 0,378 | 70 | 0,294 | 177 |
| Journal of the Royal Statistical Society, Series A | 1 | 0,001 | 629 | 0,270 | 577 | 0,000 | 593 |
| Journal of the Social Sciences | 0 | 0,000 | 673 | 0,000 | 673 | 0,000 | 593 |
| Journal of Transnational Management | 11 | 0,015 | 373 | 0,308 | 399 | 0,068 | 379 |
| Journal of Transport Economics and Policy | 20 | 0,027 | 245 | 0,350 | 199 | 0,174 | 275 |
| Journal of Urban Economics | 23 | 0,031 | 216 | 0,340 | 252 | 0,104 | 331 |
| Journal of World Business | 9 | 0,012 | 412 | 0,300 | 449 | 0,064 | 384 |
| Journal of World Trade | 1 | 0,001 | 629 | 0,255 | 609 | 0,000 | 593 |
| Kansantaloudellinen Aikakauskirja | 2 | 0,003 | 590 | 0,263 | 595 | 0,000 | 593 |
| Keio Economic Studies | 2 | 0,003 | 590 | 0,268 | 580 | 0,000 | 593 |
| Kobe Economic and Business Review | 2 | 0,003 | 590 | 0,281 | 535 | 0,000 | 589 |
| Kredit und Kapital | 0 | 0,000 | 673 | 0,000 | 673 | 0,000 | 593 |
| Kyklos | 44 | 0,059 | 65 | 0,383 | 53 | 0,217 | 238 |
| Kyoto Economic Review | 0 | 0,000 | 673 | 0,000 | 673 | 0,000 | 593 |
| Labor History | 12 | 0,016 | 357 | 0,302 | 434 | 0,053 | 408 |
| Labour | 22 | 0,030 | 224 | 0,346 | 218 | 0,815 | 33 |
| Labour Economics | 27 | 0,036 | 176 | 0,349 | 203 | 0,159 | 281 |
| LActualité Economique/Revue DAnalyse Economique | 24 | 0,032 | 208 | 0,341 | 241 | 0,349 | 147 |
| Lahore Journal of Economics | 0 | 0,000 | 673 | 0,000 | 673 | 0,000 | 593 |
| Land Economics | 5 | 0,007 | 498 | 0,290 | 484 | 0,063 | 387 |
| Law and Contemporary Problems | 0 | 0,000 | 673 | 0,000 | 673 | 0,000 | 593 |
| Lecturas de Economia | 2 | 0,003 | 590 | 0,259 | 603 | 0,000 | 593 |
| Liiketaloudellinen Aikakauskirja | 3 | 0,004 | 553 | 0,293 | 477 | 0,016 | 501 |
| LImpresa | 0 | 0,000 | 673 | 0,000 | 673 | 0,000 | 593 |
| LIndustria, Nuova Serie | 1 | 0,001 | 629 | 0,240 | 637 | 0,000 | 593 |
| Local Economy | 9 | 0,012 | 412 | 0,301 | 435 | 0,033 | 446 |
| Macroeconomic Dynamics | 68 | 0,091 | 12 | 0,398 | 19 | 1,036 | 18 |
| Management | 3 | 0,004 | 553 | 0,244 | 632 | 0,013 | 512 |
| Managerial and Decision Economics | 54 | 0,072 | 31 | 0,395 | 21 | 1,090 | 17 |
| Manchester School | 25 | 0,034 | 202 | 0,357 | 167 | 0,082 | 362 |
| Margin | 2 | 0,003 | 590 | 0,234 | 646 | 0,001 | 574 |
| Marine Resource Economics | 3 | 0,004 | 553 | 0,265 | 587 | 0,001 | 580 |
| Maritime Economics and Logistics | 9 | 0,012 | 412 | 0,311 | 383 | 0,060 | 392 |
| Maritime Policy and Management | 1 | 0,001 | 629 | 0,246 | 630 | 0,000 | 593 |
| Marketing Science | 13 | 0,017 | 335 | 0,300 | 444 | 0,311 | 167 |

| Journal | Degree | Normalized degree | Rank degree | Centrality | Rank centrality | Betweenness (x100) | Rank betweenness |
|---|---|---|---|---|---|---|---|
| Mathematical Finance | 18 | 0,024 | 271 | 0,324 | 331 | 0,113 | 321 |
| Mathematical Methods of Operations Research | 5 | 0,007 | 498 | 0,275 | 566 | 0,013 | 515 |
| Mathematical Social Sciences | 24 | 0,032 | 208 | 0,337 | 265 | 0,186 | 264 |
| Metrika | 1 | 0,001 | 629 | 0,237 | 639 | 0,000 | 593 |
| Metroeconomica | 32 | 0,043 | 136 | 0,363 | 132 | 0,238 | 224 |
| Michigan Academician | 0 | 0,000 | 673 | 0,000 | 673 | 0,000 | 593 |
| Michigan Law Review | 1 | 0,001 | 629 | 0,237 | 641 | 0,000 | 593 |
| Middle East Journal | 2 | 0,003 | 590 | 0,246 | 628 | 0,000 | 593 |
| Middle East Technical University Studies in Development | 14 | 0,019 | 319 | 0,344 | 232 | 0,047 | 420 |
| Modern Asian Studies | 4 | 0,005 | 530 | 0,271 | 575 | 0,021 | 481 |
| Momento Económico | 0 | 0,000 | 673 | 0,000 | 673 | 0,000 | 593 |
| Moneda y Crédito | 27 | 0,036 | 176 | 0,353 | 180 | 0,293 | 178 |
| Moneta e Credito | 16 | 0,021 | 288 | 0,327 | 315 | 0,377 | 136 |
| Monthly Labor Review | 1 | 0,001 | 629 | 0,255 | 609 | 0,000 | 593 |
| Multinational Finance Journal | 27 | 0,036 | 176 | 0,362 | 145 | 0,082 | 363 |
| National Institute Economic Review | 2 | 0,003 | 590 | 0,266 | 583 | 0,000 | 593 |
| National Tax Journal | 22 | 0,030 | 224 | 0,345 | 224 | 0,198 | 253 |
| Nationaløkonomisk Tidsskrift | 3 | 0,004 | 553 | 0,283 | 526 | 0,001 | 584 |
| Natural Resource Modeling | 5 | 0,007 | 498 | 0,293 | 473 | 0,010 | 526 |
| Natural Resources Journal | 0 | 0,000 | 673 | 0,000 | 673 | 0,000 | 593 |
| Netnomics | 10 | 0,013 | 393 | 0,336 | 268 | 0,022 | 480 |
| New Political Economy | 24 | 0,032 | 208 | 0,342 | 239 | 0,820 | 32 |
| New Zealand Economic Papers | 0 | 0,000 | 673 | 0,000 | 673 | 0,000 | 593 |
| New Zealand Geographer | 0 | 0,000 | 673 | 0,000 | 673 | 0,000 | 593 |
| Nigerian Journal of Economic and Social Studies | 0 | 0,000 | 673 | 0,000 | 673 | 0,000 | 593 |
| Nonlinear Dynamics, Psychology, and Life Sciences | 7 | 0,009 | 454 | 0,327 | 313 | 0,011 | 521 |
| Nonprofit Management and Leadership | 3 | 0,004 | 553 | 0,259 | 601 | 0,006 | 546 |
| Nordic Journal of Political Economy | 0 | 0,000 | 673 | 0,000 | 673 | 0,000 | 593 |
| North American Actuarial Journal | 2 | 0,003 | 590 | 0,260 | 599 | 0,000 | 593 |
| North American Journal of Economics and Finance | 42 | 0,056 | 78 | 0,379 | 65 | 0,299 | 173 |
| OECD Economic Studies | 1 | 0,001 | 629 | 0,244 | 631 | 0,000 | 593 |
| Okonomi og Politik | 1 | 0,001 | 629 | 0,242 | 635 | 0,000 | 593 |
| OPEC Review | 0 | 0,000 | 673 | 0,000 | 673 | 0,000 | 593 |
| Open Economies Review | 57 | 0,077 | 24 | 0,390 | 32 | 0,775 | 39 |
| OR Spectrum | 13 | 0,017 | 335 | 0,307 | 408 | 0,544 | 77 |
| Organization and Environment | 6 | 0,008 | 479 | 0,296 | 465 | 0,026 | 467 |
| Oxford Bulletin of Economics and Statistics | 13 | 0,017 | 335 | 0,329 | 306 | 0,016 | 503 |
| Oxford Development Studies | 55 | 0,074 | 29 | 0,390 | 28 | 0,996 | 22 |
| Oxford Economic Papers | 25 | 0,034 | 202 | 0,352 | 187 | 0,137 | 295 |
| Oxford Review of Economic Policy | 15 | 0,020 | 300 | 0,330 | 292 | 0,032 | 447 |
| Pacific Economic Bulletin | 9 | 0,012 | 412 | 0,306 | 411 | 0,008 | 536 |
| Pacific Economic Review | 124 | 0,166 | 1 | 0,449 | 1 | 3,932 | 1 |
| Pacific-Basin Finance Journal | 50 | 0,067 | 43 | 0,367 | 115 | 0,631 | 57 |
| Pakistan Development Review | 0 | 0,000 | 673 | 0,000 | 673 | 0,000 | 593 |
| Papeles de Economía Española | 0 | 0,000 | 673 | 0,000 | 673 | 0,000 | 593 |
| Papers in Regional Science | 32 | 0,043 | 136 | 0,351 | 197 | 0,505 | 83 |
| Pesquisa e Planejamento Econômico | 3 | 0,004 | 553 | 0,273 | 570 | 0,000 | 593 |
| PharmacoEconomics | 3 | 0,004 | 553 | 0,250 | 620 | 0,003 | 562 |
| Philippine Review of Economics | 5 | 0,007 | 498 | 0,310 | 388 | 0,002 | 569 |
| Policy | 2 | 0,003 | 590 | 0,234 | 647 | 0,000 | 593 |
| Policy Review | 0 | 0,000 | 673 | 0,000 | 673 | 0,000 | 593 |
| Policy Sciences | 2 | 0,003 | 590 | 0,249 | 622 | 0,020 | 486 |
| Policy Studies | 3 | 0,004 | 553 | 0,255 | 611 | 0,000 | 593 |
| Política Economica | 11 | 0,015 | 373 | 0,304 | 425 | 0,034 | 441 |
| Political Science Quarterly | 2 | 0,003 | 590 | 0,236 | 643 | 0,003 | 556 |
| Politická Ekonomie | 13 | 0,017 | 335 | 0,310 | 389 | 0,085 | 357 |
| Population | 1 | 0,001 | 629 | 0,207 | 664 | 0,000 | 593 |
| Population and Development Review | 5 | 0,007 | 498 | 0,312 | 380 | 0,660 | 52 |
| Population Bulletin | 1 | 0,001 | 629 | 0,246 | 628 | 0,000 | 593 |
| Population Research and Policy Review | 6 | 0,008 | 479 | 0,285 | 512 | 0,021 | 483 |
| Population Review | 3 | 0,004 | 553 | 0,283 | 520 | 0,071 | 376 |
| Population Studies | 1 | 0,001 | 629 | 0,207 | 664 | 0,000 | 593 |
| Post-Communist Economies | 17 | 0,023 | 280 | 0,320 | 349 | 0,326 | 156 |
| Post-Soviet Affairs | 7 | 0,009 | 454 | 0,277 | 552 | 0,127 | 305 |
| Prague Economic Papers | 13 | 0,017 | 335 | 0,310 | 389 | 0,085 | 357 |
| Problemas del Desarrollo | 0 | 0,000 | 673 | 0,000 | 673 | 0,000 | 593 |
| Problems of Economic Transition | 2 | 0,003 | 590 | 0,212 | 663 | 0,000 | 593 |
| Public Administration Review | 0 | 0,000 | 673 | 0,000 | 673 | 0,000 | 593 |
| Public Budgeting and Finance | 0 | 0,000 | 673 | 0,000 | 673 | 0,000 | 593 |
| Public Choice | 69 | 0,093 | 10 | 0,405 | 6 | 1,539 | 7 |
| Public Finance | 0 | 0,000 | 673 | 0,000 | 673 | 0,000 | 593 |
| Public Finance Review | 16 | 0,021 | 288 | 0,330 | 297 | 0,259 | 203 |
| Public Policy Research | 0 | 0,000 | 673 | 0,000 | 673 | 0,000 | 593 |
| Quaderni storici | 1 | 0,001 | 629 | 0,181 | 669 | 0,000 | 593 |
| Quantitative Finance | 41 | 0,055 | 83 | 0,382 | 57 | 0,569 | 69 |
| Quarterly Journal of Austrian Economics | 21 | 0,028 | 235 | 0,343 | 234 | 0,399 | 127 |
| Quarterly Journal of Business and Economics | 13 | 0,017 | 335 | 0,343 | 234 | 0,210 | 243 |
| Quarterly Journal of Economics | 35 | 0,047 | 119 | 0,374 | 89 | 0,282 | 188 |
| Quarterly Review of Economics and Finance | 43 | 0,058 | 72 | 0,382 | 55 | 0,535 | 78 |
| RAND Journal of Economics | 23 | 0,031 | 216 | 0,339 | 258 | 0,125 | 310 |
| Rassegna Economica | 2 | 0,003 | 590 | 0,213 | 662 | 0,000 | 593 |
| Real Estate Economics | 32 | 0,043 | 136 | 0,363 | 131 | 0,272 | 194 |
| Recherches Economiques de Louvain/Louvain Economic Review | 24 | 0,032 | 208 | 0,343 | 233 | 0,274 | 192 |
| Région et Développement | 14 | 0,019 | 319 | 0,311 | 385 | 0,221 | 235 |
| Regional Science and Urban Economics | 45 | 0,060 | 62 | 0,379 | 66 | 0,562 | 71 |
| Regional Studies | 28 | 0,038 | 165 | 0,352 | 187 | 0,557 | 72 |
| Regulation | 28 | 0,038 | 165 | 0,358 | 163 | 0,200 | 252 |

| Journal | Degree | Normalized degree | Rank degree | Centrality | Rank centrality | Betweenness (x100) | Rank betweenness |
|---|---|---|---|---|---|---|---|
| Research in Economics | 39 | 0,052 | 96 | 0,369 | 107 | 0,340 | 151 |
| Research in Law and Economics | 0 | 0,000 | 673 | 0,000 | 673 | 0,000 | 593 |
| Research Policy | 21 | 0,028 | 235 | 0,335 | 273 | 0,190 | 259 |
| Research Review | 1 | 0,001 | 629 | 0,276 | 563 | 0,000 | 593 |
| Resource and Energy Economics | 31 | 0,042 | 145 | 0,369 | 107 | 0,324 | 157 |
| Resources Policy | 17 | 0,023 | 280 | 0,333 | 277 | 0,086 | 356 |
| Review of Accounting Studies | 10 | 0,013 | 393 | 0,303 | 432 | 0,031 | 449 |
| Review of African Political Economy | 7 | 0,009 | 454 | 0,300 | 444 | 0,049 | 419 |
| Review of Agricultural Economics | 3 | 0,004 | 553 | 0,255 | 612 | 0,010 | 525 |
| Review of Austrian Economics | 19 | 0,026 | 259 | 0,326 | 321 | 0,054 | 406 |
| Review of Black Political Economy | 4 | 0,005 | 530 | 0,300 | 450 | 0,000 | 585 |
| Review of Derivatives Research | 32 | 0,043 | 136 | 0,348 | 205 | 0,173 | 276 |
| Review of Development Economics | 66 | 0,089 | 13 | 0,400 | 13 | 0,716 | 48 |
| Review of Economic Conditions in Italy | 8 | 0,011 | 433 | 0,293 | 477 | 0,029 | 461 |
| Review of Economic Design | 34 | 0,046 | 125 | 0,365 | 123 | 0,138 | 294 |
| Review of Economic Dynamics | 29 | 0,039 | 155 | 0,357 | 165 | 0,179 | 269 |
| Review of Economic Studies | 44 | 0,059 | 65 | 0,374 | 86 | 0,504 | 84 |
| Review of Economics and Statistics | 36 | 0,048 | 112 | 0,362 | 137 | 0,200 | 250 |
| Review of Finance | 40 | 0,054 | 90 | 0,365 | 124 | 0,311 | 166 |
| Review of Financial Economics | 32 | 0,043 | 136 | 0,356 | 171 | 0,258 | 205 |
| Review of Financial Studies | 18 | 0,024 | 271 | 0,317 | 362 | 0,020 | 489 |
| Review of Income and Wealth | 16 | 0,021 | 288 | 0,335 | 271 | 0,422 | 118 |
| Review of Industrial Organization | 15 | 0,020 | 300 | 0,333 | 276 | 0,223 | 234 |
| Review of International Economics | 94 | 0,126 | 2 | 0,419 | 2 | 1,514 | 8 |
| Review of International Political Economy | 29 | 0,039 | 155 | 0,343 | 236 | 0,748 | 42 |
| Review of International Studies | 12 | 0,016 | 357 | 0,290 | 487 | 0,330 | 154 |
| Review of Political Economy | 30 | 0,040 | 148 | 0,352 | 185 | 0,293 | 179 |
| Review of Quantitative Finance and Accounting | 44 | 0,059 | 65 | 0,363 | 132 | 0,499 | 86 |
| Review of Radical Political Economics | 6 | 0,008 | 479 | 0,281 | 538 | 0,012 | 518 |
| Review of Regional Studies | 21 | 0,028 | 235 | 0,314 | 371 | 0,177 | 273 |
| Review of Social Economy | 26 | 0,035 | 189 | 0,345 | 221 | 0,356 | 146 |
| Review of Urban and Regional Development Studies | 28 | 0,038 | 165 | 0,352 | 187 | 0,366 | 141 |
| Review of World Economics/Weltwirtschaftliches Archiv | 34 | 0,046 | 125 | 0,365 | 121 | 0,249 | 214 |
| Revista Brasileira de Economia | 0 | 0,000 | 673 | 0,000 | 673 | 0,000 | 593 |
| Revista de Analisis Economico | 25 | 0,034 | 202 | 0,341 | 245 | 0,422 | 119 |
| Revista de Economía | 0 | 0,000 | 673 | 0,000 | 673 | 0,000 | 593 |
| Revista de Economía Aplicada | 10 | 0,013 | 393 | 0,301 | 443 | 0,034 | 440 |
| Revista de Economía Institucional | 13 | 0,017 | 335 | 0,325 | 328 | 0,144 | 287 |
| Revista de Economía Política | 8 | 0,011 | 433 | 0,290 | 486 | 0,085 | 359 |
| Revista de Economía Política/Brazilian Journal of Political Economy | 20 | 0,027 | 245 | 0,335 | 275 | 0,446 | 103 |
| Revista de Economía y Estadística, N.S. | 2 | 0,003 | 590 | 0,250 | 621 | 0,000 | 593 |
| Revista de Estudios Políticos | 2 | 0,003 | 590 | 0,246 | 627 | 0,000 | 593 |
| Revista de Historia Económica | 9 | 0,012 | 412 | 0,287 | 498 | 0,062 | 388 |
| Revista de Historia Industrial | 7 | 0,009 | 454 | 0,277 | 551 | 0,034 | 443 |
| Revue Canadienne des Sciences de lAdministration/Canadian Journal of Administrative | 4 | 0,005 | 530 | 0,291 | 482 | 0,008 | 535 |
| Revue de LOFCE | 3 | 0,004 | 553 | 0,260 | 597 | 0,107 | 326 |
| Revue déconomie du Développement | 2 | 0,003 | 590 | 0,257 | 608 | 0,000 | 593 |
| Revue déconomie Financière | 9 | 0,012 | 412 | 0,300 | 451 | 0,014 | 508 |
| Revue déconomie Industrielle | 16 | 0,021 | 288 | 0,317 | 362 | 0,179 | 268 |
| Revue déconomie Politique | 22 | 0,030 | 224 | 0,340 | 253 | 0,315 | 162 |
| Revue déconomie Regionale et Urbaine | 14 | 0,019 | 319 | 0,320 | 344 | 0,200 | 251 |
| Revue dEtudes Comparatives Est-Ouest | 3 | 0,004 | 553 | 0,278 | 548 | 0,007 | 541 |
| Revue Economique | 20 | 0,027 | 245 | 0,336 | 270 | 0,202 | 249 |
| Revue Finance Contrôle Stratégie | 5 | 0,007 | 498 | 0,282 | 533 | 0,259 | 202 |
| Revue Française de Gestion | 1 | 0,001 | 629 | 0,215 | 659 | 0,000 | 593 |
| Revue Française déconomie | 13 | 0,017 | 335 | 0,315 | 369 | 0,136 | 297 |
| Revue Tiers Monde | 4 | 0,005 | 530 | 0,275 | 565 | 0,012 | 517 |
| RISEC: International Review of Economics and Business | 42 | 0,056 | 78 | 0,380 | 63 | 0,600 | 61 |
| Risk Decision and Policy | 31 | 0,042 | 145 | 0,364 | 126 | 0,143 | 288 |
| Rivista di Politica Economica | 26 | 0,035 | 189 | 0,362 | 139 | 0,384 | 133 |
| Rivista di Storia Economica, N.S. | 9 | 0,012 | 412 | 0,305 | 417 | 0,020 | 491 |
| Rivista Internazionale di Scienze Sociali | 6 | 0,008 | 479 | 0,304 | 425 | 0,030 | 454 |
| Rivista Italiana degli Economisti | 11 | 0,015 | 373 | 0,328 | 312 | 0,023 | 478 |
| Scandinavian Economic History Review | 2 | 0,003 | 590 | 0,252 | 616 | 0,000 | 586 |
| Scandinavian Journal of Economics | 17 | 0,023 | 280 | 0,337 | 267 | 0,061 | 389 |
| Schmollers Jahrbuch: Zeitschrift für Wirtschafts- und Sozialwissenschaften/Journal of | 5 | 0,007 | 498 | 0,276 | 560 | 0,013 | 516 |
| Schweizerische Zeitschrift für Volkswirtschaft und Statistik/Swiss Journal of Economics | 0 | 0,000 | 673 | 0,000 | 673 | 0,000 | 593 |
| Schweizerische Zeitschrift für Wirtschafts- und Finanzmarktrecht / Revue suisse de dro | 0 | 0,000 | 673 | 0,000 | 673 | 0,000 | 593 |
| Science and Society | 8 | 0,011 | 433 | 0,303 | 433 | 0,140 | 292 |
| Scottish Journal of Political Economy | 16 | 0,021 | 288 | 0,347 | 212 | 0,025 | 471 |
| Seoul Journal of Economics | 28 | 0,038 | 165 | 0,364 | 128 | 0,091 | 352 |
| Singapore Economic Review | 48 | 0,064 | 47 | 0,376 | 77 | 0,376 | 137 |
| Sloan Management Review | 14 | 0,019 | 319 | 0,332 | 284 | 0,135 | 298 |
| Small Business Economics | 23 | 0,031 | 216 | 0,360 | 158 | 0,400 | 126 |
| Social and Economic Studies | 2 | 0,003 | 590 | 0,254 | 613 | 0,007 | 542 |
| Social Choice and Welfare | 57 | 0,077 | 24 | 0,391 | 26 | 0,523 | 81 |
| Social Research | 0 | 0,000 | 673 | 0,000 | 673 | 0,000 | 593 |
| Social Science Japan Journal | 8 | 0,011 | 433 | 0,301 | 442 | 0,058 | 398 |
| Social Science Quarterly | 0 | 0,000 | 673 | 0,000 | 673 | 0,000 | 593 |
| Social Security Bulletin | 0 | 0,000 | 673 | 0,000 | 673 | 0,000 | 593 |
| Society | 4 | 0,005 | 530 | 0,282 | 528 | 0,077 | 368 |
| South African Journal of Economic and Management Sciences, N.S. | 5 | 0,007 | 498 | 0,301 | 435 | 0,008 | 533 |
| South African Journal of Economics | 19 | 0,026 | 259 | 0,350 | 200 | 0,147 | 285 |
| Southern Economic Journal | 0 | 0,000 | 673 | 0,000 | 673 | 0,000 | 593 |
| Soviet Studies | 10 | 0,013 | 393 | 0,319 | 353 | 0,101 | 338 |
| Spanish Economic Review | 10 | 0,013 | 393 | 0,303 | 430 | 0,014 | 507 |
| Spoudai | 1 | 0,001 | 629 | 0,237 | 638 | 0,000 | 593 |

| Journal | Degree | Normalized degree | Rank degree | Centrality | Rank centrality | Betweenness (x100) | Rank betweenness |
|---|---|---|---|---|---|---|---|
| Statistica | 4 | 0,005 | 530 | 0,237 | 641 | 0,491 | 90 |
| Statistica Applicata | 2 | 0,003 | 590 | 0,188 | 667 | 0,241 | 221 |
| Statistical Journal | 0 | 0,000 | 673 | 0,000 | 673 | 0,000 | 593 |
| Statistical Papers | 14 | 0,019 | 319 | 0,322 | 340 | 0,679 | 51 |
| Strategic Finance | 0 | 0,000 | 673 | 0,000 | 673 | 0,000 | 593 |
| Structural Change and Economic Dynamics | 65 | 0,087 | 14 | 0,404 | 7 | 1,621 | 6 |
| Studi Economici | 5 | 0,007 | 498 | 0,272 | 574 | 0,197 | 254 |
| Studies in Economics and Finance | 13 | 0,017 | 335 | 0,321 | 342 | 0,243 | 220 |
| Studies in Family Planning | 5 | 0,007 | 498 | 0,269 | 579 | 0,489 | 91 |
| Studies in Nonlinear Dynamics and Econometrics | 42 | 0,056 | 78 | 0,374 | 86 | 0,223 | 233 |
| Supreme Court Economic Review | 3 | 0,004 | 553 | 0,283 | 524 | 0,027 | 464 |
| Survey of Current Business | 0 | 0,000 | 673 | 0,000 | 673 | 0,000 | 593 |
| Swiss Political Science Review | 9 | 0,012 | 412 | 0,277 | 556 | 0,020 | 490 |
| Tahqiqat-e eqtesadi (Quarterly Journal of Economic Research) | 0 | 0,000 | 673 | 0,000 | 673 | 0,000 | 593 |
| Teaching Business and Economics | 0 | 0,000 | 673 | 0,000 | 673 | 0,000 | 593 |
| Technology Analysis and Strategic Management | 5 | 0,007 | 498 | 0,277 | 559 | 0,013 | 514 |
| Technology and Culture | 7 | 0,009 | 454 | 0,291 | 481 | 0,044 | 427 |
| Telecommunications Policy | 11 | 0,015 | 373 | 0,318 | 355 | 0,239 | 223 |
| Theory and Decision | 42 | 0,056 | 78 | 0,381 | 58 | 0,646 | 55 |
| Tijdschrift voor Economie en Management | 8 | 0,011 | 433 | 0,314 | 372 | 0,028 | 463 |
| Tourism and Hospitality Management | 1 | 0,001 | 629 | 0,186 | 668 | 0,000 | 593 |
| Tourism Economics | 3 | 0,004 | 553 | 0,235 | 644 | 0,245 | 216 |
| Transnational Corporations | 15 | 0,020 | 300 | 0,337 | 265 | 0,226 | 230 |
| Transportation | 12 | 0,016 | 357 | 0,304 | 428 | 0,052 | 416 |
| Transportation Journal | 0 | 0,000 | 673 | 0,000 | 673 | 0,000 | 593 |
| Transportation Research: Part A: Policy and Practice | 13 | 0,017 | 335 | 0,316 | 366 | 0,107 | 325 |
| Transportation Research: Part B: Methodological | 19 | 0,026 | 259 | 0,330 | 295 | 0,272 | 193 |
| Transportation Research: Part D: Transport and Environment | 12 | 0,016 | 357 | 0,316 | 367 | 0,056 | 401 |
| Transportation Research: Part E: Logistics and Transporation Review | 15 | 0,020 | 300 | 0,339 | 255 | 0,356 | 145 |
| Travail et Emploi | 2 | 0,003 | 590 | 0,293 | 473 | 0,000 | 593 |
| Ukrainian Economic Review | 1 | 0,001 | 629 | 0,230 | 650 | 0,000 | 593 |
| UN Chronicle | 0 | 0,000 | 673 | 0,000 | 673 | 0,000 | 593 |
| Urban Studies | 8 | 0,011 | 433 | 0,287 | 502 | 0,018 | 497 |
| Venture Capital | 7 | 0,009 | 454 | 0,307 | 407 | 0,020 | 487 |
| Vierteljahrsheftze zur Wirtschaftsforschung | 7 | 0,009 | 454 | 0,310 | 392 | 0,042 | 428 |
| Water Resources Research | 3 | 0,004 | 553 | 0,243 | 634 | 0,052 | 413 |
| Wirtschaftspolitische Blätter | 0 | 0,000 | 673 | 0,000 | 673 | 0,000 | 593 |
| WorkingUSA | 13 | 0,017 | 335 | 0,313 | 378 | 0,132 | 302 |
| World Bank Economic Review | 48 | 0,064 | 47 | 0,388 | 37 | 0,442 | 106 |
| World Bank Research Observer | 13 | 0,017 | 335 | 0,336 | 269 | 0,210 | 242 |
| World Development | 43 | 0,058 | 72 | 0,375 | 82 | 1,030 | 19 |
| World Economy | 73 | 0,098 | 5 | 0,403 | 8 | 1,156 | 14 |
| Yale Journal on Regulation | 1 | 0,001 | 629 | 0,214 | 661 | 0,000 | 593 |
| Yale Law Journal | 4 | 0,005 | 530 | 0,281 | 535 | 0,245 | 217 |
| Yapi Kredi Economic Review | 1 | 0,001 | 629 | 0,234 | 645 | 0,000 | 593 |
| Zagreb International Review of Economics and Business | 3 | 0,004 | 553 | 0,287 | 501 | 0,001 | 579 |
| Zbornik Radova Ekonomskog Fakulteta U. Rijeci/Proceedings of Rijeka School of Econc | 2 | 0,003 | 590 | 0,230 | 649 | 0,000 | 593 |
| Zeitschrift für ArbeitsmarktForschung/Journal for Labour Market Research | 3 | 0,004 | 553 | 0,277 | 556 | 0,007 | 539 |
| Zeitschrift für Betriebswirtschaft | 6 | 0,008 | 479 | 0,274 | 568 | 0,052 | 415 |
| Zeitschrift für Wirtschaftspolitik | 1 | 0,001 | 629 | 0,278 | 548 | 0,000 | 593 |